\documentclass[10pt]{iopart}

\usepackage{graphicx}
\usepackage{amsfonts}
\expandafter\let\csname equation*\endcsname\relax
\expandafter\let\csname endequation*\endcsname\relax
\usepackage{amsmath}
\usepackage{enumerate}
\usepackage{comment}
\usepackage{color}
\usepackage{citesort}
\newcommand{\bA}{\mathbf A }
\newcommand{\R}{{\rm I\!R}}
\newcommand{\N}{{\rm I\!N}}
\newcommand{\bra}{\langle}
\newcommand{\ket}{\rangle}
\newcommand{\balpha}{\boldsymbol\alpha}
\newcommand{\bn}{\boldsymbol n}

\newcommand{\D}{\textit{D}}

\newcommand{\order}{{\cal O}}

\begin{document}

\title[Exactly Solvable Random Graph Ensemble with Extensively Many Short Cycles]{Exactly Solvable Random Graph Ensemble with Extensively Many Short Cycles}

\author{Fabi\'an Aguirre L\'opez$^{1,2}$, Paolo Barucca$^{3,4}$, Mathilde Fekom$^5$, and Anthony CC Coolen$^{1,2}$}
\address{$^1$Department of Mathematics, King's College London, The Strand, London WC2R 2LS, United Kingdom\\ 
$^2$Institute for Mathematical and Molecular Biomedicine, King's College London, Hodgkin Building, London SE1 1UL, United Kingdom\\
$^3$London Institute for Mathematical Sciences, 35a South St, Mayfair, London W1K 2XF, United Kingdom\\
$^4$University of Z\"{u}rich, Department of Banking and Finance, Z\"urich, ZH, Switzerland\\
$^5$UFR de Physique, Universit\'e Paris Diderot (Paris 7), 5 Rue Thomas Mann, 75013 Paris, France
}

\ead{fabian.aguirre\_lopez@kcl.ac.uk, paolo.barucca@bf.uzh.ch, mathilde.fekom1@yahoo.fr, ton.coolen@kcl.ac.uk}

\begin{abstract}
We introduce and analyse ensembles of 2-regular random graphs with a tuneable distribution of short cycles. The phenomenology of these graphs depends critically on the scaling of the ensembles' control parameters relative to the number of nodes. A phase diagram is presented, showing a second order phase transition from a connected to a disconnected phase. We study both the canonical formulation, where the size is large but fixed, and the grand canonical formulation, where the size is sampled from a discrete distribution, and show their equivalence in the thermodynamical limit. 
We also compute analytically  the spectral density, which consists of a discrete set of isolated eigenvalues, representing short cycles, and a continuous part, representing cycles of diverging size.
\end{abstract}

\maketitle


\section{Introduction}

Models with pairwise interacting elements are ubiquitous in physics and are sufficient to capture the phenomenology of many  systems, ranging from condensed matter via biology to the social sciences and informatics. The properties of the network of interactions strongly affects the properties of the system under study, and hence the analysis of networks is central in modern physics.
Testing statistically whether a specific network property influences the dynamics of a system requires sampling networks where such a property is controllable, and for which the probability measure is known.
This is why random graph ensembles, especially maximum entropy ones from which it is possible to sample networks systematically with  controlled properties, have gained increasing popularity. They range from degree configuration models \cite{bekessy1972asymptotic,bender1978asymptotic,molloy1995critical,newman2001random,catanzaro2005generation,annibale2017generating}, i.e. where the number of connections or nodes are fixed, to more complicated models where a block structure is given to the nodes in the network, as in stochastic block models \cite{holland1983stochastic}, or other ensembles where clustering, i.e. the tendency of nodes with common neighbours to be connected, is enhanced \cite{jonasson1999random,holme2002growing,davidsen2002emergence,burda2004network,krapivsky2005network,newman2009random,bollobas2011sparse,bianconi2014triadic}.
However, controlling analytically and numerically second or higher-order properties of networks, i.e. node properties that not only depend on first neighbours, such as the density of cycles, is still a great mathematical and analytical challenge, whose range of applications continues to grow \cite{granovetter1973strength,davidsen2002emergence,sole2002model,vazquez2003growing,marsili2004rise,ispolatov2005duplication,toivonen2006model,jackson2007meeting}.

So far, nearly all analytical results obtained for random graph ensembles rely on the assumption of the absence of short cycles, the \emph{tree-like approximation}, and we have analytical solutions only for random graphs were clustering is absent or too weak (or improbable) to be relevant.
One of the first random graph ensembles in literature to include short cycles was \cite{strauss1986general}, where a term depending on the number of 3-cycles of the graph was included as a modification to the well known Erd\"os-R\'enyi model (ER). This was done in order to encourage this connection transitivity in the graph. 
However, as was found in simulations \cite{strauss1986general} and in a more rigorous way in \cite{jonasson1999random,burda2004network}, unless the graph is particularly small, this approach does not allow for a tuneable number of triangles. 
Depending on the values and the scaling of the parameters, the model of \cite{strauss1986general}  either stays in a phase very close to the ER model, with a very slight increase in triangles, or it collapses to a condensed phase, where the complete clique has probability one. This abrupt transition was found to be a generic feature of  exponential random graph models. As was shown in \cite{chatterjee2013estimating,yin2016detailed}, this phenomenon will be observed  not only in two-parameter models like the Strauss model, but in any exponential graph ensemble  that is biased such as to induce a finite number of subgraph densities.

The natural way to prevent clique formation in the condensed phase is to study random graph ensembles with hard degree constraints.  Here all graphs have exactly the same degree distribution, and this distribution is chosen such that the complete clique is not an allowed state. However, this constraint makes analytical solution intractable, leaving numerical sampling from the ensemble as the only route for investigation. Examples are the Poisonnian graphs studied numerically in \cite{avetisov2016eigenvalue}, where it was found that a triangle bias induced finite size graphs to  break down into small clusters to maximize the triangle density. Regular graphs with triangle bias were numerically explored in \cite{annibale2017generating}, and showed similar phenomenology. However, both Poissonnian and regular graph enembles with triangle bias have so far resisted analytical solution.

In this paper we introduce an exactly solvable ensemble of 2-regular random graphs, with an exponential measure that controls the presence of short cycles up to any finite length. The imposition of 2-regularity removes the possibility of a complete clique forming, and forces the graph instead to be partitioned into a set of disconnected cycles of different lengths. This makes the ensemble analytically solvable and perfectly tuneable. The model displays a second-order transition, from a phase dominated by extensively long cycles, to a phase where only (extensively many) cycles of short lengths are present.

In section 2 we introduce and solve the model, in its canonical formulation. In section 3 we describe analytically  the phases of the ensemble and the critical hyper-surface in the space of parameters; from this result we also compute analytically  the spectral density of the ensemble.
In section 4 we demonstrate the equivalence of the canonical and grand canonical formulations of the model, and in section 5 we show the agreement of the analytical predictions with numerical experiments. 
In a  final discussion section we summarize the results and delineate the future directions for this research, which are twofold.  The first is to  relax the 2-regularity constraint of the ensemble, in order to make it more directly comparable to realistic networks. The second is to understand better recent analytical approaches to random graph ensembles that involve  constraints on the number of closed paths of all lengths, which is equivalent to constraining random graphs via their spectra  \cite{CoolenLoopy}.

\section{Definitions}

We define a random graph ensemble over the set of undirected simple regular graphs of degree 2, which we denote by $\mathcal{G}_N$. Any graph in $\mathcal{G}_N$ is necessarily a set of disjoint cycles. The probability assigned to each graph $\mathbf A \in \mathcal{G}_N$ is chosen proportional to the exponential of a weighted sum of the number of  triangles, squares, pentagons, $\ldots$ , K-cycles present in $\bA$. We refer to this as biasing with respect of the number of short cycles. Thus
\begin{eqnarray}
\label{CD:prob}
    p(\mathbf A)&=\frac{1}{Z_N(\boldsymbol\alpha)}\exp \left(\sum_{\ell=3}^K \ell \alpha_\ell n_\ell(\mathbf A)\right),
  \end{eqnarray}
 Here $n_\ell(\mathbf A)$ denotes the number of length-$\ell$ cycles, i.e. closed paths of length $\ell$ without backtracking and without over-counting, and $\boldsymbol\alpha= (\alpha_3,\dots,\alpha_K)\in\R^{K-2}$ is a vector of control parameters.  Note that isolated nodes ($\ell=1$) and dimers ($\ell=2$) cannot occur due to the degree constraint. The factors $\ell$ in (\ref{CD:prob})  are included for later convenience. We are effectively biasing with respect to the total number of $\ell$-cycles starting at a given node through the introduction of the field $\alpha_\ell$. 
 
The partition function $Z_N(\boldsymbol\alpha)$ is  given by
\begin{eqnarray}
    \label{partitionFunctionCD}
    Z_N(\boldsymbol\alpha)&=\sum_{\mathbf A\in \mathcal{G}_N}\exp\left(\sum_{\ell=3}^K \ell \alpha_\ell n_\ell(\mathbf A)\right).
\end{eqnarray}
Expression (\ref{CD:prob}) defines a maximum entropy random graph ensemble with respect to the $K-2$ observables $n_\ell(\mathbf A)$, 
whose ensemble averages are controlled by varying  the parameters $\boldsymbol\alpha$. We choose $K$ to be a fixed number for all values of $N$. This exponential form is a particular version of the one presented in equation (1.1) of \cite{chatterjee2013estimating}. It is an ensemble where we are interested in controlling the expected values of a finite number of graph observables.

The average fraction of the $N$ nodes that will be found in an $\ell$-cycle is given by  
\begin{eqnarray}
    \label{CD:densityLoop_l}
    m_\ell&=& \frac{\ell}{N}\bra n_\ell(\bA)\ket.
\end{eqnarray}
where $\bra f(\bA)\ket=\sum_{\bA}p(\bA)f(\bA)$.
Following the statistical mechanics route,  we define a generating function $\phi_N(\boldsymbol\alpha)$:
\begin{eqnarray}
    \label{CD:phiDef}
    \phi_N(\boldsymbol\alpha)&=& N^{-1}\log [Z_N(\boldsymbol\alpha)/N!].
\end{eqnarray}
The main quantities of interest (\ref{CD:densityLoop_l}) for our graph ensemble (\ref{CD:prob}) can be computed from (\ref{CD:phiDef}) via
\begin{eqnarray}
    \label{CD:densityDerivative}
    m_\ell&=& \partial\phi_N(\boldsymbol\alpha)/\partial\alpha_\ell.
\end{eqnarray}

The generator $\phi_N(\balpha)$ is minus the free energy density, apart from a
factor $N!$ which reflects (topologically irrelevant) node label permutations. Including this factor will ensure that 
the limit $\phi(\balpha)=\lim_{N\to\infty}\phi_N(\balpha)$ exists.

\section{Analytical solution}

\subsection{Summation over graphs}

To evaluate the partition function (\ref{partitionFunctionCD}) we need to perform a sum over graphs. Such sums are usually not analytically tractable, especially when the ensemble definition involves cycles, as is the case in (\ref{CD:prob}). Here we are able to perform the summation by rewriting it as
\begin{eqnarray}
    Z_N (\balpha)= \sum_{\bn}\D(\bn) \rme^{\sum_{\ell=3}^K \ell\alpha_\ell n_\ell},
\end{eqnarray}
with $\bn=(n_3,\ldots,n_N)\in\N^{N-2}$. 
This decomposition reflects the fact that, in the particular case of $\mathcal{G}_N$, we are fortunate that each graph has to be a collection of cycles, and can therefore be identified fully by a sequence $\bn=(n_3,\dots,n_N)$ that specifies the number of cycles of each possible length up to $N$, and a labelling of the nodes. The sum over graphs is then performed by summing over all possible sequences $\bn$, keeping track of the multiplicity of each sequence via an associated density of states $\D(\bn)$:
\begin{eqnarray}
\D(\bn)&=& \sum_{\mathbf A\in \mathcal{G}_N}\prod_{\ell=3}^N \delta_{n_\ell,n_\ell(\mathbf{ A})}
=\frac{N! ~\delta_{N,\sum_{\ell=3}^N \ell n_\ell}}{\prod_{\ell=3}^N[ (2\ell)^{n_\ell}n_\ell !]}.~~
\end{eqnarray}
 Apart from the condition $N=\sum_{\ell=3}^N \ell n_\ell$, this density 
 is proportional to $N!$ but corrected for over-counting due to the indistinguishability of different length-$\ell$ cycles, 
giving a divisor $n_\ell !$, and due to the different ways one can number the nodes in each $\ell$-cycle without altering the graph ($\ell$ cyclic permutations, plus $\ell$ anti-cyclic permutations), giving a further divisor $(2\ell)^{n_\ell}$. Using the integral form of the Kronecker delta $\delta_{nm}=\int_{-\pi}^{\pi}(\rmd\omega/2\pi)\rme^{\rmi \omega(n-m)}$, we can thus write the partition function as
\begin{eqnarray}
\hspace*{-10mm}
    Z_N(\boldsymbol\alpha)&=&\sum_{\bn}\frac{N!}{\prod_{\ell=3}^N [(2\ell)^{n_\ell}n_\ell!]}
    \left(\prod_{\ell=3}^K \rme^{\ell \alpha_\ell n_\ell}\right)
    \int_{-\pi}^{\pi}\frac{\rmd\omega}{2\pi}\rme^{\rmi\omega\left(N-\sum_{\ell=3}^N \ell n_\ell\right)}
     \nonumber
   \\
   \hspace*{-10mm}
    &=& \frac{N!}{2\pi}\int_{-\pi}^{\pi}\rmd\omega~ \rme^{\rmi \omega N}\prod_{\ell=3}^K\left(\sum_{n_\ell\geq 0}\frac{\rme^{(\alpha_\ell-\rmi \omega)\ell n_\ell}}{(2\ell)^{n_\ell}n_\ell!}\right)
 \prod_{\ell=K+1}^N\! \left(\sum_{n_\ell\geq 0}\frac{\rme^{-\rmi \omega \ell n_\ell}}{(2\ell)^{n_\ell}n_\ell!}\right)\nonumber
   \\
   \hspace*{-10mm}
    &=& \frac{N!}{2\pi}\int_{-\pi}^{\pi}\rmd\omega~\exp
    \left(\rmi \omega N+\sum_{\ell=3}^K\frac{\rme^{(\alpha_\ell-\rmi \omega)\ell }}{2\ell}
    +\sum_{\ell=K+1}^N\! \frac{\rme^{-\rmi \omega \ell}}{2\ell}\right).
     \label{CD:calcZ}
\end{eqnarray}
From this, in combination with (\ref{CD:phiDef}), we infer that 
\begin{eqnarray}
    \phi_N(\boldsymbol\alpha)&=& \frac{1}{N}\log 
    \int_{-\pi}^{\pi}\frac{\rmd\omega}{2\pi}~\rme^{Nf_N(\omega,\balpha)}
    \label{eq:phi_saddle}
    \end{eqnarray}
    with 
    \begin{eqnarray}
    f_N(\omega,\balpha)&=& 
    \rmi \omega +\sum_{\ell=3}^K\frac{\rme^{(\alpha_\ell-\rmi \omega)\ell }}{2\ell N}
    +\sum_{\ell=K+1}^N\! \frac{\rme^{-\rmi \omega \ell}}{2\ell N}.
    \label{eq:f_N}
\end{eqnarray}
An exact expression for (\ref{eq:phi_saddle}), valid for any finite $N$, would require to perform the integral in it. Instead, we proceed in the usual way as in statistical physics. We look at the thermodynamic limit, focusing then on $\phi(\balpha)=\lim_{N\to\infty}\phi_N(\balpha)$. This will allow us to calculate the asymptotic expressions for (\ref{CD:densityLoop_l}), which should differ from the finite size values by $\order(1/N)$ corrections.

The limit $N\to\infty$ of (\ref{eq:phi_saddle}) can now be obtained by evaluating the integral over $\omega$ in (\ref{CD:calcZ}) via steepest descent:
\begin{eqnarray}
    \label{CD:phi}
    \phi(\boldsymbol\alpha)=\lim_{N\to\infty}{\rm extr}_{\omega}f_N(\omega,\boldsymbol\alpha).
\end{eqnarray}
The extremum is found by solving $\partial f(\omega,\boldsymbol\alpha)/\partial\omega=0$.

\subsection{Scaling with $N$ of control parameters}

We observe that for finite $\{\alpha_\ell\}$
our model  cannot exhibit nonzero cycle densities $m_\ell$ in the infinite size limit, since the $\boldsymbol\alpha$-dependent term  in (\ref{eq:f_N}) vanishes for $N\to\infty$. We are therefore led to redefining the parameters $\balpha$ with a size dependent shift, 
\begin{eqnarray}
    \alpha_\ell&=& \tilde\alpha_\ell+\frac{1}{\ell}\log(N),
\end{eqnarray}
where $\tilde{\alpha}_\ell=\order(1)$. An intuitive explanation for this scaling is presented in section \ref{section:GCD}. 
We denote the vector of shifted $\order(1)$ control parameters by $ \tilde{\boldsymbol\alpha} =(\tilde\alpha_3,\dots,\tilde\alpha_K)$, and we define 
$ \phi_N(\boldsymbol\alpha)=\varphi_N(\tilde{\boldsymbol\alpha})$. This implies that for  $N\to\infty$ we will have  $m_\ell=\partial  \varphi(\tilde\balpha)/\partial\tilde{\alpha}_\ell$, in which now
\begin{eqnarray}
 \varphi(\tilde{\boldsymbol\alpha})&=&
 \lim_{N\to\infty}{\rm extr}_\omega\Big\{
 \textrm i \omega +\sum_{\ell=3}^K \frac{\rme^{(\tilde{\alpha}_\ell-\rmi \omega) \ell}}{2\ell}
    +\sum_{\ell=K+1}^N\frac{\rme^{-\rmi \omega \ell}}{2\ell N}
    \Big\}.
\end{eqnarray}
Differentiation of this latter expression reveals that 
the value $\omega_N$ at the extremum is to be solved from 
\begin{eqnarray}
 1  &=& \frac{1}{2}\sum_{\ell=3}^K \rme^{(\tilde{\alpha}_\ell-\rmi \omega_N) \ell}
    +\frac{1}{2N}\sum_{\ell=K+1}^N \rme^{-\rmi \omega_N \ell},
    \label{eq:omega_problem}
\end{eqnarray}
and that the asymptotic values of the observables $m_\ell$ are  subsequently given by
\begin{eqnarray}
m_\ell=\frac{1}{2}\rme^{(\tilde\alpha_\ell-\rmi\omega_N)\ell}.
\label{eq:m_expression}
\end{eqnarray}
This last identity, in combination with (\ref{eq:omega_problem}), prompts us to introduce $m_\infty=1-\sum_{\ell\leq K} m_\ell\in[0,1]$, which gives the fraction of the nodes that are {\em not} in cycles of length $K$ or less. It is for $N\to\infty$ apparently given by
\begin{eqnarray}
m_\infty&=& \lim_{N\to\infty}\frac{1}{2N}\sum_{\ell=K+1}^N \rme^{-\rmi \omega_N \ell}.
\end{eqnarray}
It follows from (\ref{eq:m_expression}),  that the physical saddle point $\omega$, after contour deformation, must be purely imaginary. We switch accordingly to the new variable $x=\rme^{-\rmi\omega}\in\R_0^+$, in terms of which our equations become:
\begin{eqnarray}
 1  &=& \frac{1}{2}\sum_{\ell=3}^K x_N^\ell \rme^{\ell \tilde{\alpha}_\ell}
    +\frac{1}{2N}\sum_{\ell=K+1}^N x_N^\ell,
    \label{eq:SP_x}
 \\
 m_\ell&=&  \lim_{N\to\infty}\frac{1}{2}x_N^\ell \rme^{\ell \tilde\alpha_\ell},
  \label{eq:SP_m}
\\
m_\infty&=& \lim_{N\to\infty}\frac{1}{2N}\sum_{\ell=K+1}^N x_N^\ell .
 \label{eq:SP_m_infty}
\end{eqnarray}

\subsection{Phase phenomenology of the ensemble}

We will now  demonstrate that the solutions to the coupled equations (\ref{eq:SP_x},\ref{eq:SP_m}) give rise to two phases of our graph ensemble. A \emph{disconnected} phase is characterized by the fact that all nodes are typically in cycles of length $K$ or less, so $m_\infty=0$. A second phase, the \emph{connected} phase, is characterized by finding a finite fraction of the nodes in longer cycles, so here $m_\infty>0$. The transition separating the phases is marked by bifurcation of  $m_\infty>0$ solutions. 

If $\lim_{N\to\infty}x_N = x<1$, the second term of (\ref{eq:SP_x}) vanishes for $N\to\infty$, and we immediately obtain $m_\infty=0$. Hence we are in the disconnected phase, and here the asymptotic observables $m_\ell$ are simply found by solving 
\begin{eqnarray}
m_\infty\!=\!0:&~~~&  1  = \frac{1}{2}\sum_{\ell=3}^K x^\ell \rme^{\ell \tilde{\alpha}_\ell},~~~
 m_\ell= \frac{1}{2}x^\ell \rme^{\ell \tilde\alpha_\ell}.
 \label{eq:D_phase}
 \end{eqnarray}
The condition $x<1$ for this solution to exist will be met for large values of $\{\tilde\alpha_\ell\}$. Upon reducing the control parameters $\{\tilde\alpha_\ell\}$, the value of $x$ will increase, and a transition to the connected phase occurs exactly when $x=1$. This happens at the critical manifold in the $K\!-\!2$ dimensional parameter space, defined by validity of
\begin{eqnarray}
\sum_{\ell=3}^K  \rme^{\ell \tilde{\alpha}_\ell}=2.
  \label{CD:criticalManifold}
 \end{eqnarray}
 
To confirm the equations of the connected phase, we need to investigate how the solution $x_N$ of (\ref{eq:SP_x}) scales with $N$ as we approach $x=1$. Substituting $x_N=1-\xi/N$, expanding (\ref{eq:SP_x})  in $N$, and taking the limit $N\to\infty$ gives
\begin{eqnarray}
 m_\ell = \frac{1}{2}\rme^{\ell \tilde\alpha_\ell},~~~~~~
m_\infty= 1- \frac{1}{2}\sum_{\ell=3}^K \rme^{\ell \tilde\alpha_\ell},
\label{eq:C_phase}
\end{eqnarray}
and the link between $\xi$ and $m_\infty$ is 
$m_\infty=(1\!-\!\rme^{-\xi})/2\xi$. 

It turns out that all cycles of finite length $L>K$ will always have vanishing density for $N\to\infty$. This can be seen  simply by replacing $K\to L$ in the previous analysis, but with $\alpha_{\ell}=0$ for all $K<\ell\leq L$. The newly added control parameters with $\ell>K$ will give $\tilde\alpha_{\ell}=-\ell^{-1}\log N$, and hence $m_\ell=\lim_{N\to\infty}\frac{1}{2}x^\ell \rme^{\ell\tilde{\alpha}_\ell}=\lim_{N\to\infty}\frac{1}{2}x^\ell /N=0$, in both phases. 
We  knew that in the disconnected phase all nodes will typically be in the controlled short cycles of length $K$ or less.
We may now conclude that,  in the connected phase, those nodes that are not in the controlled short cycles (the fraction $m_\infty>0$) will typically be found in cycles of {\em diverging} length. 

The ensemble's Shannon entropy \cite{infotheory} is given by
\begin{eqnarray}
S_N&=& -\sum_{\bA\in \mathcal{G}_N}p(\mathbf A)\log p(\mathbf A)
\nonumber
\\
&=& \log N!+  N\Big[\phi_N(\boldsymbol\alpha)
-\sum_{\ell=3}^K  \alpha_\ell \frac{\partial\phi_N(\boldsymbol\alpha)}{\partial\alpha_\ell}\Big]
\nonumber
\\
&=& N\log(N)\Big(1-\sum_{\ell=3}^K  \frac{m_\ell}{\ell}\Big)+\order(N).
\end{eqnarray}
Since $\sum_{\ell=3}^K  (m_\ell/\ell)\leq \frac{1}{3}\sum_{\ell=3}^K  m_\ell\leq \frac{1}{3}$, the leading order will for large $N$ always scale as $N\log(N)$, and be bounded according to $S_N\geq \frac{2}{3}N\log (N)+\order(N)$, but with a reduced prefactor if we increase the fraction of nodes in short cycles. The lower bound is achieved in the disconnected phase, when $m_3=1$ and $m_{\ell>3}=0$.

\subsection{Spectral densities of adjacency matrices}

A graph can be represented uniquely by its adjacency matrix $\{A_{ij}\}$, where $A_{ij}\in \{0,1\}$, and $A_{ij}=1$ if and only if there is a link from $j$ to $i$. The set  $\mathcal{G}_N$ contains only simple nondirected graphs, so our  adjacency matrices are symmetric and with zero diagonal elements. The eigenvalue density of the adjacency  matrix of a graph $\bA$, 
\begin{eqnarray}
    \varrho(\mu|\mathbf A)&=&\frac{1}{N}\sum_{i=1}^N \delta[\mu-\mu_i(\bA)],
\end{eqnarray}
contains valuable information on the statistics  of cycles in the graph. Here the sum runs over the set of (real) eigenvalues $\{\mu_i(\bA)\}_{i=1,\dots,N}$ of $\mathbf A$, taking into account multiplicities. For instance, the number of closed paths in $\bA$ is proportional to $\int\!\rmd\mu~\varrho(\mu|\bA)\mu^\ell$. Our main quantity of interest will be the expected density, averaged over the ensemble probabilities (\ref{CD:prob}), in the infinite size limit, 
\begin{eqnarray}
    \varrho(\mu)=\lim_{N\rightarrow\infty}\sum_{\mathbf A \in \mathcal{G}_N}p(\mathbf A)\varrho(\mu|\mathbf A).
\end{eqnarray}

The adjacency matrix of a graph that consists of a single cycle of length $\ell$ has the Toeplitz form, and is therefore diagonalized trivially, leading to the density
\begin{eqnarray}
    \label{CD:spectrumLoop}
    \varrho_\ell (\mu)&=&\frac{1}{\ell}\sum_{r=0}^{\ell-1}\delta\Big(\mu-2\cos(2\pi r/\ell)\Big).
    \label{eq:spec_ell}
\end{eqnarray}
If the cycle length $\ell$ diverges, this density becomes continuous (in a distributional sense), see e.g. \cite{mckay1981expected}, 
\begin{eqnarray}
    \varrho_{\infty}(\mu)&=&\lim_{\ell\rightarrow\infty}\varrho_\ell(\mu)=\frac{1}{\pi}\frac{\theta(2-|\mu|)}{\sqrt{4-\mu^2}}.
     \label{eq:spec_infty}
\end{eqnarray}
 The set of eigenvalues for each $\bA\in \mathcal{G}_N$ will just be the union of all the sets of eigenvalues of the disjoint cycles of which it is composed, taking multiplicities into account:
\begin{eqnarray}
    \varrho(\mu|\mathbf A)&=&\frac{1}{N}\sum_{\ell=3}^{N}n_\ell (\mathbf A)\sum_{r=0}^{\ell-1}\delta\Big(\mu-2\cos\Big(\frac{2\pi r}{\ell})\Big)
 \nonumber   \\
    &=&\sum_{\ell=3}^N\frac{\ell n_\ell(\mathbf A)}{N}\varrho_\ell(\mu).
\end{eqnarray}
Upon averaging over the ensemble, using  (\ref{CD:densityLoop_l}) and our earlier observation that for $N\to\infty$  the fraction of nodes in cycles of finite length $L>K$ vanishes, we immediately obtain the asymptotic ensemble-averaged spectrum corresponding to (\ref{CD:prob}), expressed in terms of (\ref{eq:spec_ell}) and (\ref{eq:spec_infty}):
\begin{eqnarray}
    \varrho(\mu)&=&\sum_{\ell=3}^K m_\ell \varrho_\ell(\mu)+m_\infty \varrho_\infty(\mu).
    \label{CD:spectra}
\end{eqnarray}
Since we are working with regular graphs, we can immediately recover the spectrum of the Laplacian operator ($\mathbf{L}=2\mathbf{I}-\bA$) by the change of variable $\mu\to 2-\lambda$.

\section{\label{section:GCD}Grand Canonical approach}

Within the canonical approach one finds that, if $N$ is sufficiently large, graphs generated randomly from (\ref{CD:prob}) will all display the same values of the main intensive  quantities, such as the  fraction of $\ell$-cycles (modulo finite size fluctuations). We expect a similar claim to hold if we sample randomly both the graphs and the number $N$ of nodes, i.e. if we work with 
 grand canonical graph ensembles. The grand partition function of our ensemble with weights $w_N=\rme^{-\mu N}/N!$ 
(where $\mu>0$) is given by
\begin{eqnarray}
    \label{GCD:grandSum}
    Q(\balpha) = \sum_{N=1}^{\infty} w_N Z_N(\balpha),
\end{eqnarray}
with $Z_N(\balpha)$ defined in (\ref{partitionFunctionCD}). The divisor $N!$ in $w_N$ will simplify our calculation, without losing the benefits of the thermodynamic limit
 (since we will find that for $\mu\to 0$ the expected system size still diverges). Direct calculation of $Q(\balpha)$ now  circumvents the integration over $\omega$: 
 \begin{eqnarray}
  Q(\balpha) &=&
  \sum_{\bn}\Big(\prod_{\ell=3}^\infty\frac{ \rme^{-\mu \ell n_\ell}}{(2\ell)^{n_\ell}n_\ell!}\Big)
    \left(\prod_{\ell=3}^K \rme^{\ell \alpha_\ell n_\ell}\right)
    \nonumber
    \\
    &=&\left[\prod_{\ell=3}^K\left(\sum_{n\geq 0}\frac{ \rme^{( \alpha_\ell -\mu) \ell n}}{(2\ell)^{n}n!}\right)\right]\!
    \left[\prod_{\ell>K}\left(\sum_{n\geq 0}
 \frac{ \rme^{-\mu \ell n}}{(2\ell)^{n}n!}\right)\right]
  \nonumber
    \\
    &=&\exp\left(\sum_{\ell=3}^K\frac{ \rme^{( \alpha_\ell -\mu) \ell}}{2\ell}
    +\sum_{\ell>K}
 \frac{ \rme^{-\mu \ell}}{2\ell}\right)
  \nonumber
    \\
    &=&\exp\left(\sum_{\ell=3}^K\frac{ \rme^{( \alpha_\ell -\mu) \ell}}{2\ell}
  -\frac{1}{2}\log(1\!-\!\rme^{-\mu})  
  -\sum_{\ell=1}^{K}
 \frac{ \rme^{-\mu \ell}}{2\ell}\right),
 \nonumber
 \\[-1mm]&&
   \end{eqnarray}
   where we used $ \sum_{\ell>0}x^\ell/\ell=-\log(1\!-\!x)$. 
From $Q(\balpha)$ we obtain, in turn, the grand potential $\Omega(\balpha) =-\log Q(\balpha) $:
\begin{eqnarray}
\Omega(\balpha) &=& 
\sum_{\ell=1}^{K}\frac{ \rme^{-\mu \ell}}{2\ell}
 -\sum_{\ell=3}^K\frac{ \rme^{( \alpha_\ell -\mu) \ell}}{2\ell}
  +\frac{1}{2}\log(1-\rme^{-\mu}).
 \end{eqnarray}
Its partial derivatives with respect to $\mu$ and $\balpha$ yield the average system size, via $\langle N \rangle=\partial\Omega(\balpha)/\partial\mu$, and the average number of length-$\ell$ cycles (for $\ell=3,\ldots ,K$), via 
$\bra n_\ell(\bA)\ket=-\ell^{-1}\partial\Omega(\balpha)/\partial\alpha_\ell$. We thereby find that
\begin{eqnarray}
\bra N\ket&=& \frac{1}{2}
\frac{\rme^{-\mu}}{1\!-\!\rme^{-\mu}}
 + \frac{1}{2}\sum_{\ell=3}^K \rme^{( \alpha_\ell -\mu) \ell}
- \frac{1}{2}\sum_{\ell=1}^{K}\rme^{-\mu \ell}
 \nonumber
 \\
 &=& 
 \frac{1}{2}
\frac{\rme^{-\mu(K+1)}}{1\!-\!\rme^{-\mu}}
 + \frac{1}{2}\sum_{\ell=3}^K \rme^{( \alpha_\ell -\mu) \ell}
 \label{eq:N_grand}
 \end{eqnarray}
and
\begin{eqnarray}
\label{eq:m_ell_GC}
\bra n_\ell(\bA)\ket &=& \frac{1}{2\ell}\rme^{( \alpha_\ell -\mu) \ell}.
\end{eqnarray}
Clearly, 
$\bra N\ket$ diverges for $\mu\to 0$, which gives our thermodynamic limit. In this limit we can then work out for $\ell\in\{3,\ldots,K\}$ the 
 ratios 
 \begin{eqnarray}
 \lim_{\mu\downarrow 0}\frac{\ell\bra n_\ell \ket}{\bra N\ket}&=& 
 \lim_{\mu\downarrow 0}\frac{\rme^{( \alpha_\ell -\mu) \ell}}
 {
\frac{\rme^{-\mu(K+1)}}{1-\rme^{-\mu}}
 + \sum_{\ell^\prime=3}^K \rme^{( \alpha_{\ell^\prime} -\mu) \ell^\prime}}=0.~~~
 \end{eqnarray}
Similar to the canonical case, any $\mu$-independent $\balpha$ will asymptotically always yield a vanishing fraction of nodes in cycles of length $\ell\leq K$. It is clear from (\ref{eq:m_ell_GC}) that without a re-parametrization, the expected value of $\ell$-cycles only increases exponentially with $\alpha_\ell$ in the thermodynamic limit. We need to re-parametrize in such a way that the expected number of $\ell$-cycles increases as the expected system size increases. The re-parametrization required to obtain a non-trivial thermodynamic limit is $\alpha_\ell = \tilde{\alpha}_\ell + \ell^{-1} \log  \bra N \ket$.
Upon following this prescription, we then reproduce the canonical result
 \begin{eqnarray}
\frac{\ell\bra n_\ell \ket}{\bra N\ket}&=&  \frac{1}{2}\rme^{( \tilde{\alpha}_\ell -\mu) \ell},
 \label{eq:grand_m}
\end{eqnarray}
and our expression (\ref{eq:N_grand}) for $\bra N\ket$ now becomes
\begin{eqnarray}
\bra N\ket&=&
\frac{\rme^{-\mu(K+1)}}{(1\!-\!\rme^{-\mu})\big(2-\sum_{\ell=3}^K \rme^{(\tilde{\alpha}_\ell -\mu) \ell}\big)}.
\label{eq:N_grand_new}
 \end{eqnarray}
The re-parametrization of $\balpha$ now depends on $\balpha$ itself, via $\langle N \rangle$, and has to be consistent with a nonnegative value for (\ref{eq:N_grand_new}), i.e.  with $\frac{1}{2}\sum_{\ell=3}^K \rme^{(\tilde{\alpha}_\ell -\mu) \ell}\leq 1$. Expression (\ref{eq:grand_m}) gives us the physical interpretation $\sum_{\ell=3}^K \ell\bra n_\ell \ket/ \bra N\ket \leq 1$. 
 In the limit $\mu\downarrow 0$ the condition becomes
$\sum_{\ell=3}^K\rme^{\ell\tilde{\alpha}_\ell}\leq 2$. In the case of inequality we have $\sum_{\ell=3}^K \ell\bra n_\ell \ket/\bra N\ket<1$, so we are in the connected phase.  The case of equality reproduces our earlier phase transition condition (\ref{CD:criticalManifold}) and we enter the disconnected phase; here the thermodynamic limit is reached already for nonzero $\mu$, and we can again recover our canonical equations, with $\exp(-\mu)$ now playing the role of the canonical order parameter $x$. 

\begin{figure}[t]
\begin{picture}(372,156)
  \put(100,0){\includegraphics[width=240\unitlength]{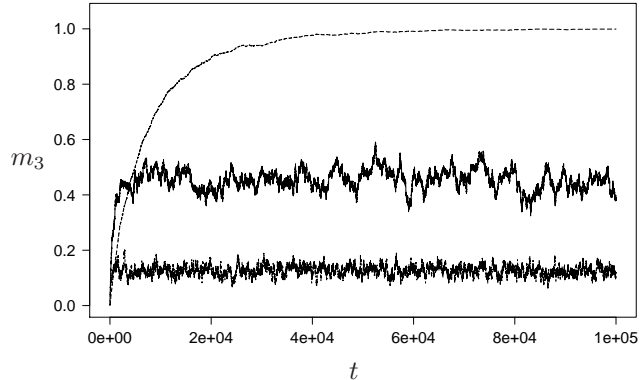}}
  \put(92,85){$m_3$}
  \put(220,4){$t$}
\end{picture}
\vspace*{-5mm}

\caption{\label{fig:Iterations}
Examples of the evolution of the fraction $m_3$ of nodes in triangles, measured during MCMC simulations, for $K=3$. Time is defined as the number of accepted edge swap moves per link. 
The bottom two curves correspond to  the connected phase of the ensemble, equilibrating to the values $m_3=0.125$ for $\tilde{\alpha}_3=\frac{1}{3}\log(0.25)$, and to $m_3=0.45$ for  $\tilde{\alpha}_3
=\frac{1}{3}\log(0.9)$. The top curve corresponds to the disconnected phase, here the MCMC process is equilibrating to the value $m_3=1$.}
\end{figure}

\begin{figure}[t]
\hspace*{1mm}
\begin{picture}(367,130)
  \put(100,0){\includegraphics[width=190\unitlength]{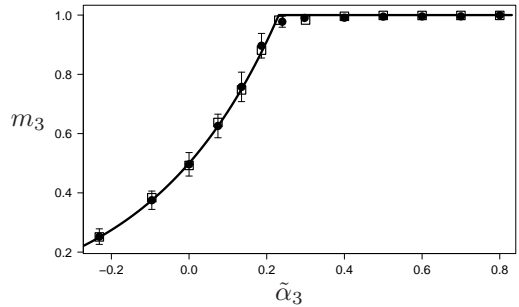}}
  \put(90,70){$m_3$}
  \put(190,4){$\tilde\alpha_3$}
\end{picture}
\vspace*{-6mm}

\caption{\label{fig:plotM3} Values of $m_3$  shown versus  $\tilde\alpha_3$ for ensembles with $K=3$. 
Numerical results, measured upon equilibration of the MCMC processes, are shown as black dots with error bars for $N=1000$, and as squares for $N=5000$ (error bars for $N=5000$ are not shown; their sizes are similar to or smaller than the squares). The solid line is the prediction of (\ref{eq:K3}). 
}
\end{figure}

\section{Numerical simulations}

\begin{figure*}[t]
\hspace*{2mm}
\begin{picture}(364,120)
  \put(0,0){\includegraphics[width=180\unitlength]{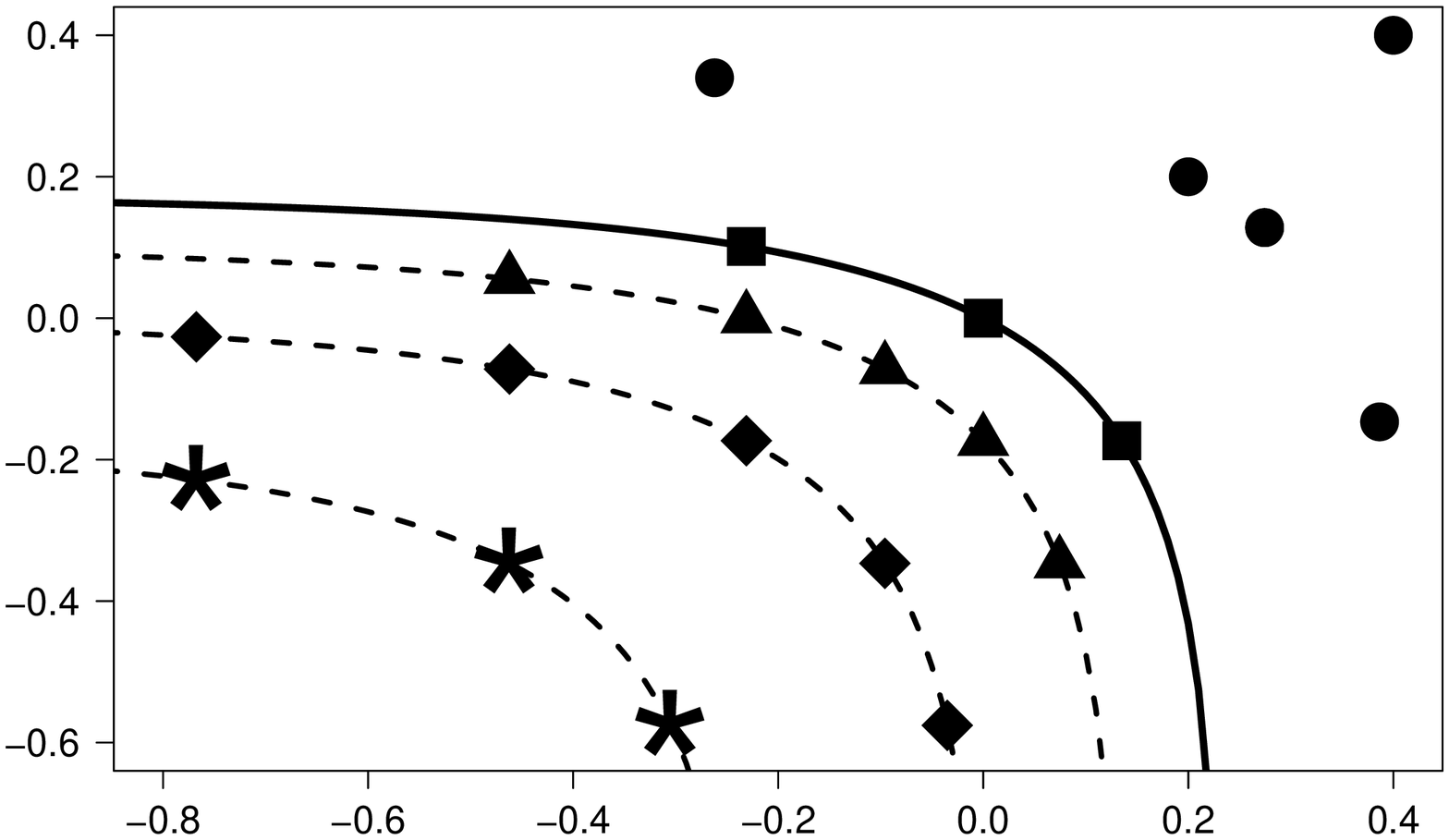}}
  \put(90,2){$\tilde\alpha_3$}  \put(-13,70){$\tilde\alpha_4$}
  \put(190,0){\includegraphics[width=180\unitlength]{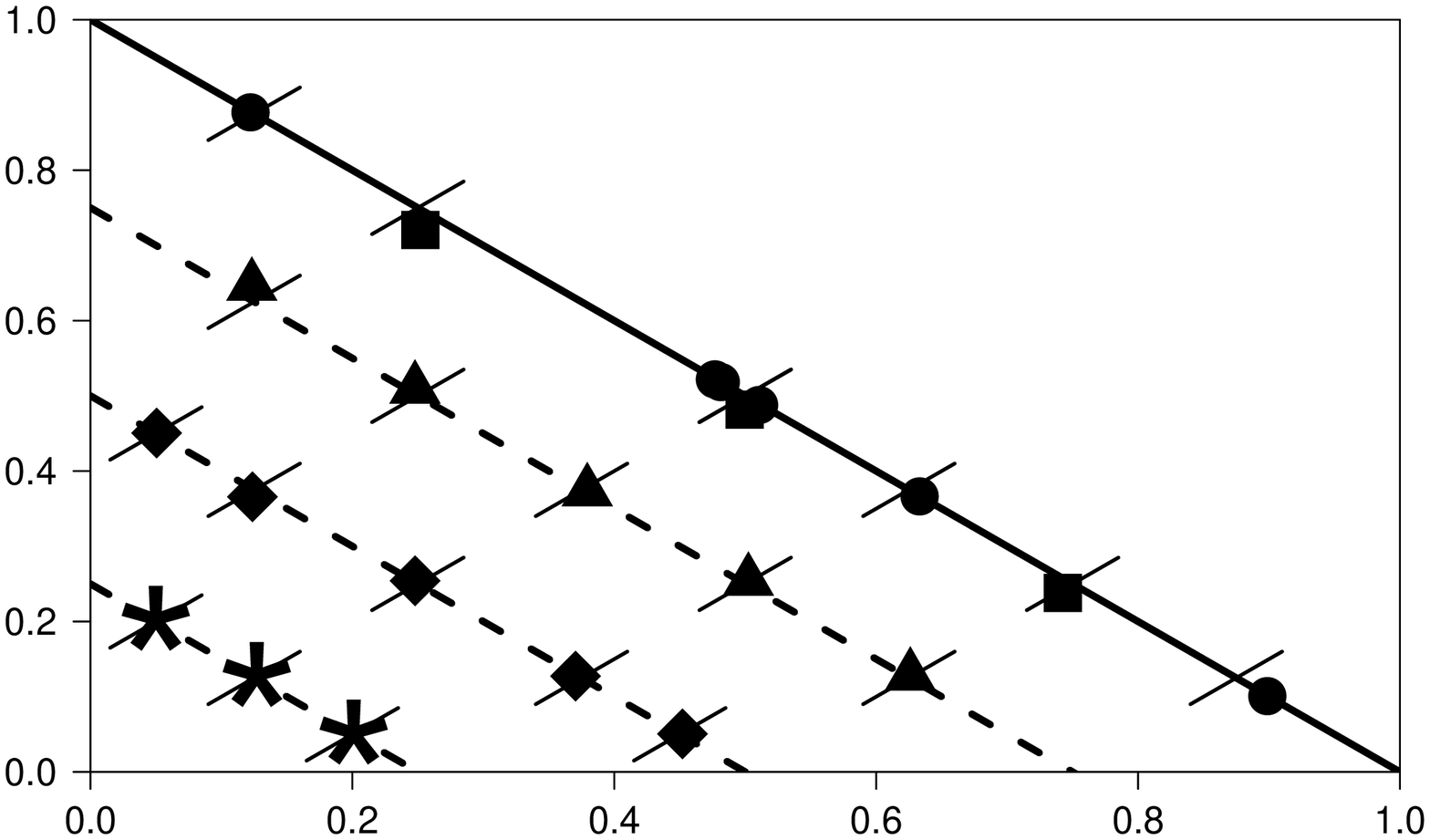}}
   \put(280,4){$m_3$} 
   \put(180,70){$m_4$}
      \put(100,100){\em Disconnected}
   \put(23,25){\em Connected}
\end{picture}
\vspace*{-5mm}

\caption{\label{fig:plotM4} Left panel: the plane of control parameters for  $K=4$. The solid black line is the critical line  
 $\rme^{3\tilde{\alpha}_3}+\rme^{4\tilde{\alpha}_4}=2$ (here $m_\infty=0$). The dashed lines correspond to parameter combinations with constant $m_\infty$, taking the values  $m_\infty\in \{0.75, 0.5, 0.25\}$, from bottom to top. The markers represent parameter combinations chosen for MCMC simulations. 
 Right panel:  the fractions $(m_3,m_4)$  associated with the control  parameter combinations in the left panel. Here the markers represent the simulation results, measured after execution of $10^4$ accepted moves per node in the MCMC to secure equilibration. The results are indeed found on the respective  lines predicted by the theory. Note that the theory predicts that all parameter combinations in the disconnected phase  $\rme^{3\tilde{\alpha}_3}+\rme^{4\tilde{\alpha}_4}\geq 2$, should be mapped to the line $m_1+m_2=1$ in the right panel. 
 Error bars were omitted, as they are as big as or smaller than the markers.}
\end{figure*}

Calculating $Z_N(\balpha)$ by numerical enumeration for nontrivial values of $N$  is not a realistic option, since the size of the set  $\mathcal{G}_N$ grows super-exponentially with $N$. Instead, to test our theoretical predictions we have sampled graphs from the ensemble (\ref{CD:prob}) using the Markov Chain Monte Carlo (MCMC) method described in e.g. \cite{sampling} or \cite{annibale2017generating}. Starting from an arbitrary 2-regular $N$-node graph, this stochastic process is based on executing repeated (degree-preserving) edge swap moves with appropriate nontrivial move acceptance probabilities, constructed such that the Markov chain's equilibrium distribution is the target measure (\ref{CD:prob}). 
In each simulation experiment, the MCMC process was first run for $10^5$ to $10^6$ accepted moves per link, and equilibration was confirmed by  measuring the Hamming distance between the instantaneous and the initial state.  After this randomization stage, the instantaneous state $\bA$ arrived at by the chain was defined to be our graph sample. We have limited our simulations to ensembles with $K=3$ and $K=4$.
The degree of equilibration achieved by the MCMC during a run of $10^5$ accepted moves per link is illustrated in  Figure \ref{fig:Iterations}, where we show typical evolution curves of the order parameter $m_3$ during the stochastic process.

For $K=3$ we have just one control parameter $\tilde\alpha_3$, and the order parameter is the fraction $m_3$ of nodes in triangles. The theory claims that, for large $N$, the graphs from our ensemble will be collections of triangles and large rings. The key equations (\ref{eq:D_phase},\ref{CD:criticalManifold},\ref{eq:C_phase}) reduce to the following predictions, with $\tilde{\alpha}_c=\frac{1}{3}\log(2)\approx 0.23105\dots$:
\begin{eqnarray}
\begin{array}{lll}
\tilde{\alpha}_3< \tilde{\alpha}_c: &~~ m_3=\frac{1}{2}\rme^{3\tilde{\alpha}_3}, &~~ {connected~phase},
\\[3mm] 
\tilde{\alpha}_3> \tilde{\alpha}_c: &~~ m_3=1, &~~ {disconnected~phase}.~~~
\end{array}
\label{eq:K3}
\end{eqnarray}
Numerical simulations with sizes $N=1000$ and $N=5000$  show excellent agreement  with these predictions, as shown in Figure \ref{fig:plotM3}, both in terms of the values of $m_3$ and in terms of the location of the transition.

\begin{figure*}[t]
\vspace*{-14mm}
\hspace*{4mm}
\begin{picture}(358,288)
  \put(5,4){\includegraphics[width=0.32\textwidth,height=50mm]{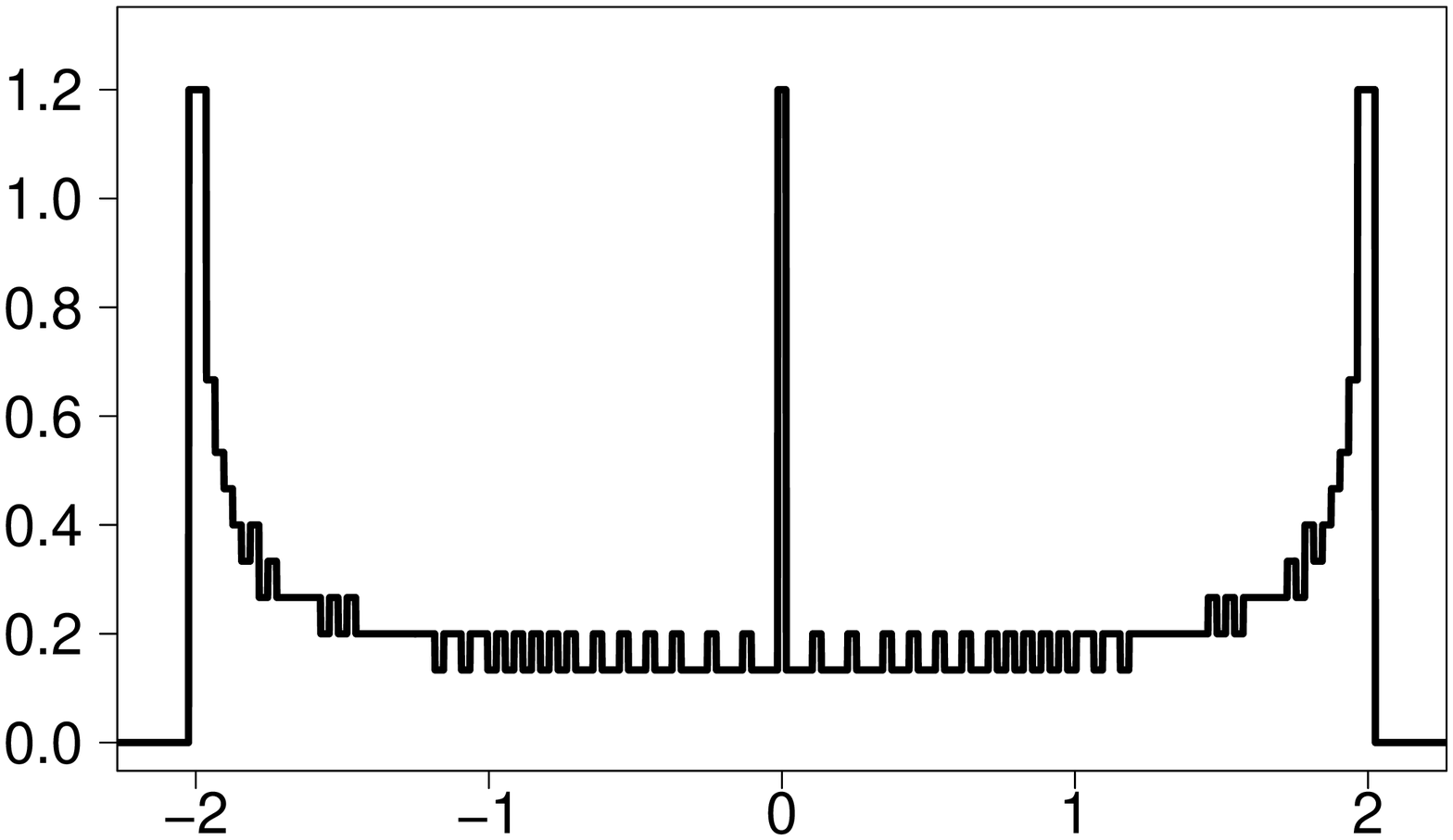}}
  \put(125,4){\includegraphics[width=0.32\textwidth,height=50mm]{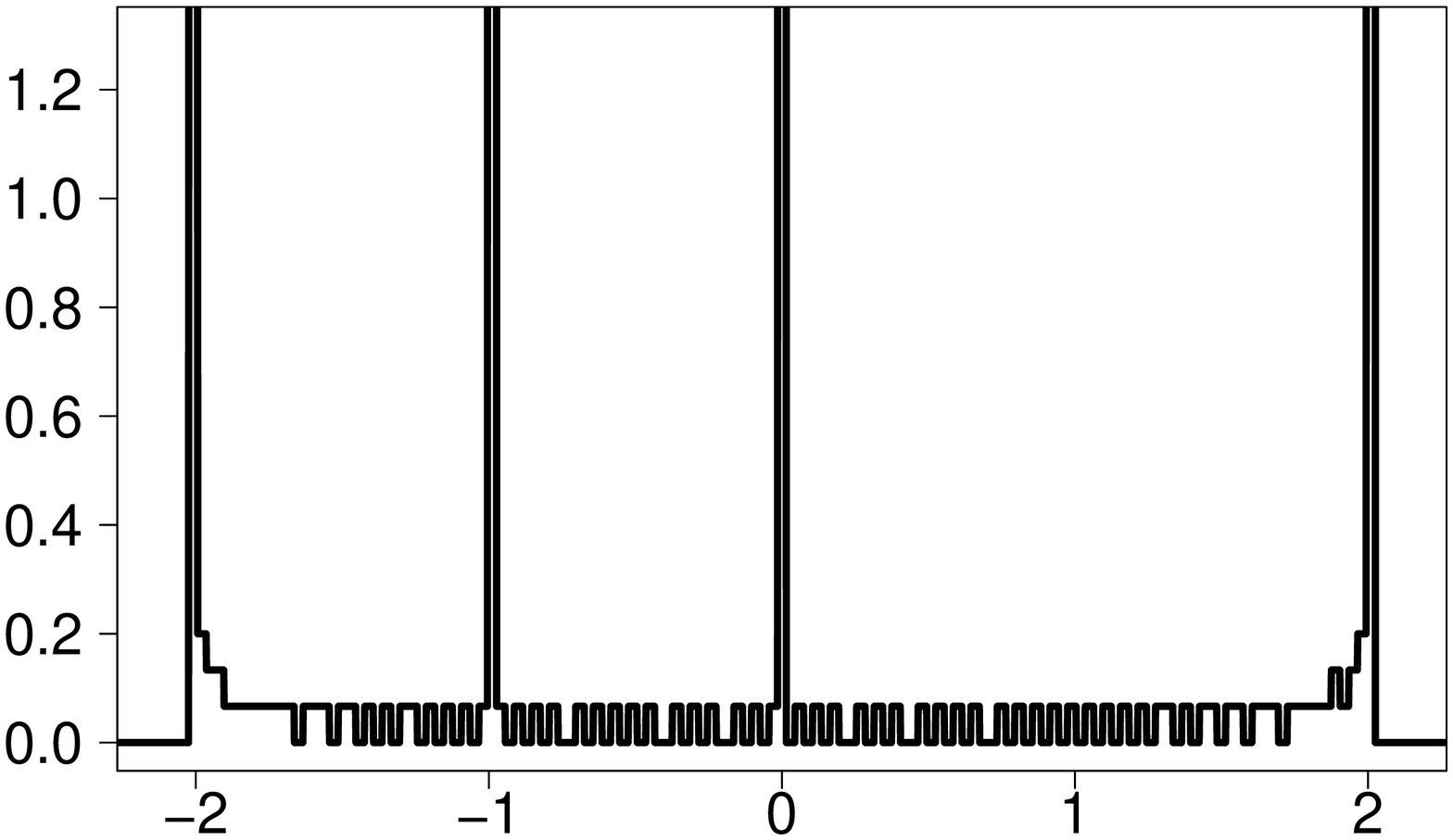}}
  \put(240,4){\includegraphics[width=0.32\textwidth,height=50mm]{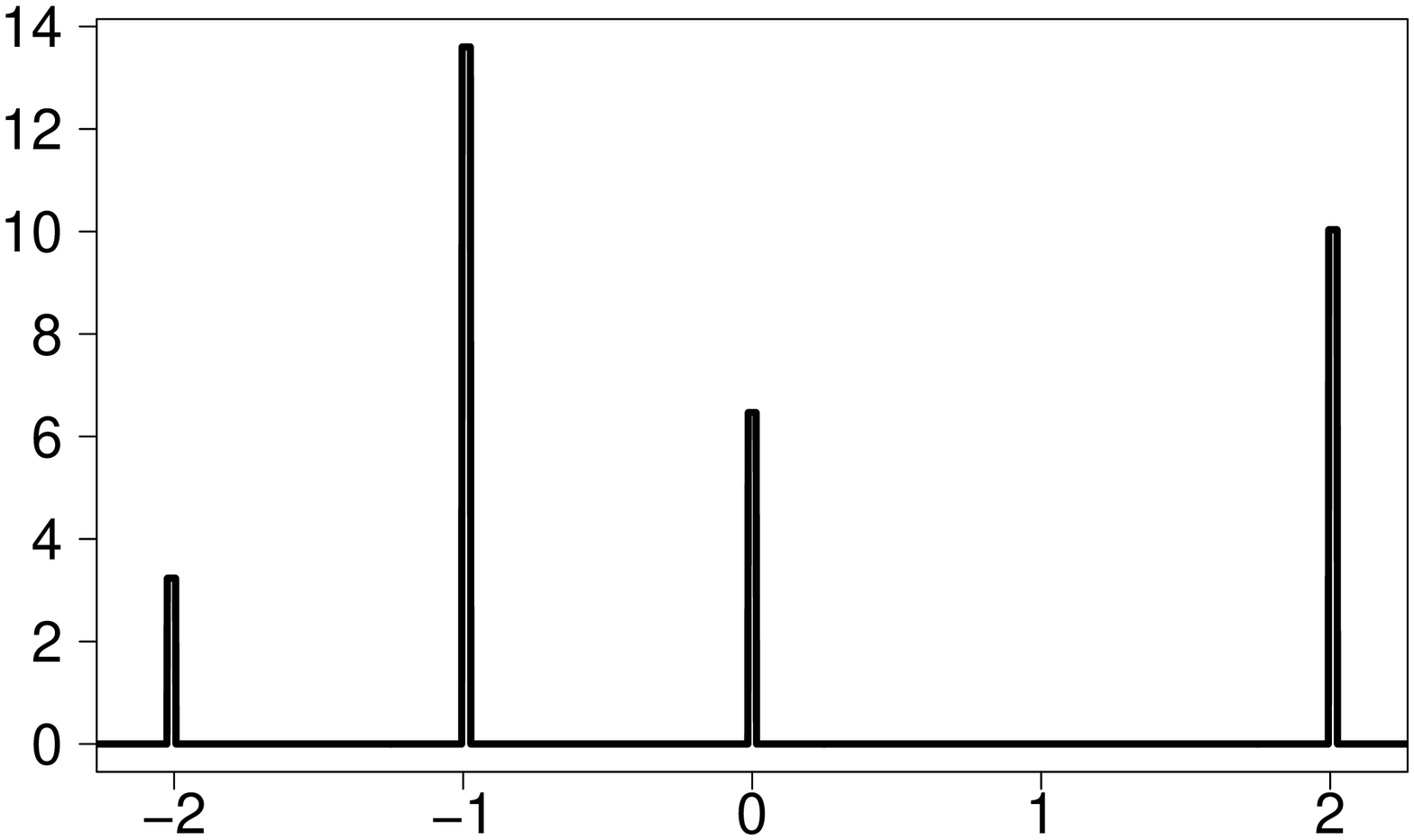}}
  \put(5,140){\includegraphics[width=0.32\textwidth,trim={0 0 0 0},clip]{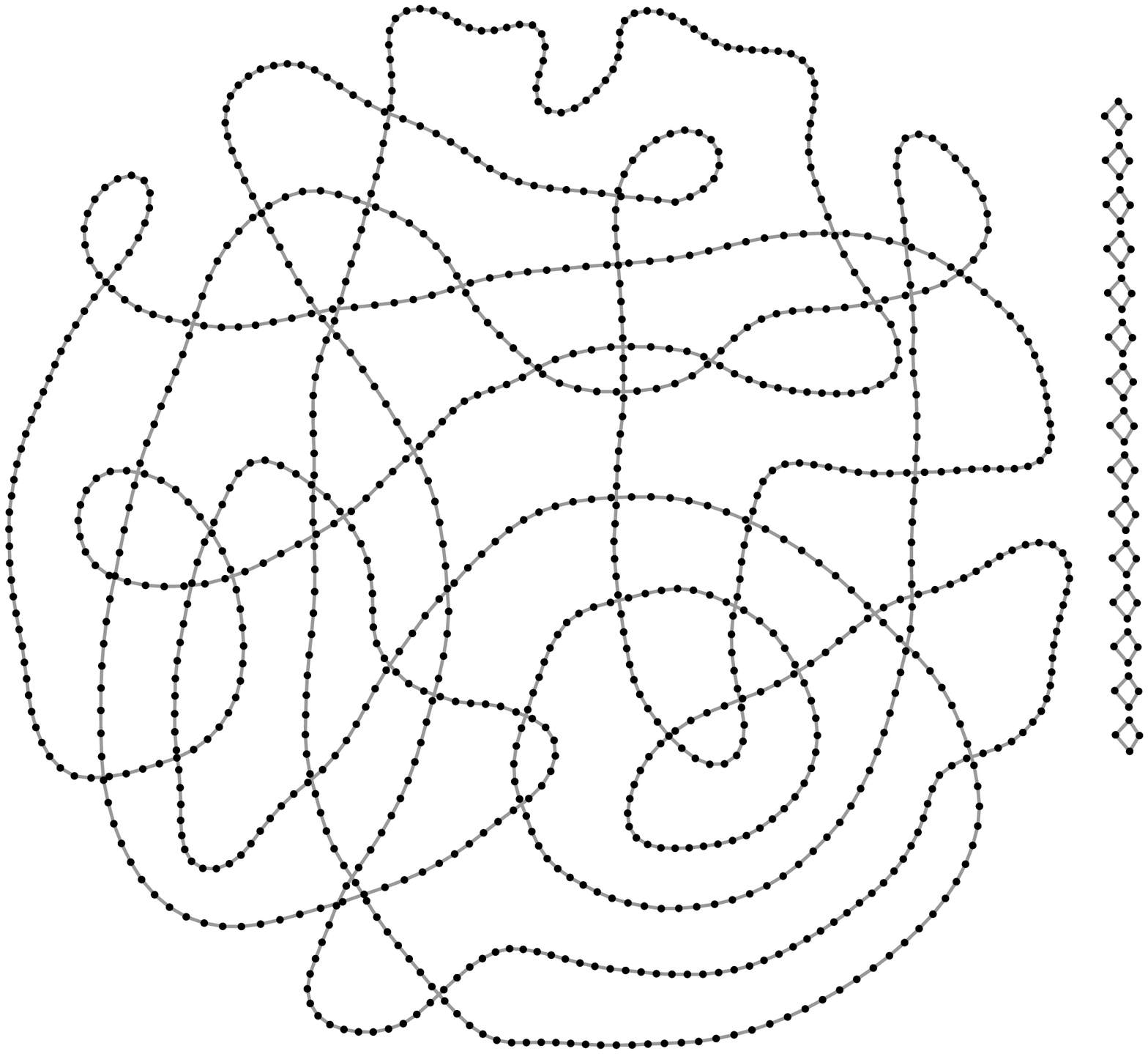}}
  \put(120,140){\includegraphics[width=0.32\textwidth,trim={0 0 0 0},clip]{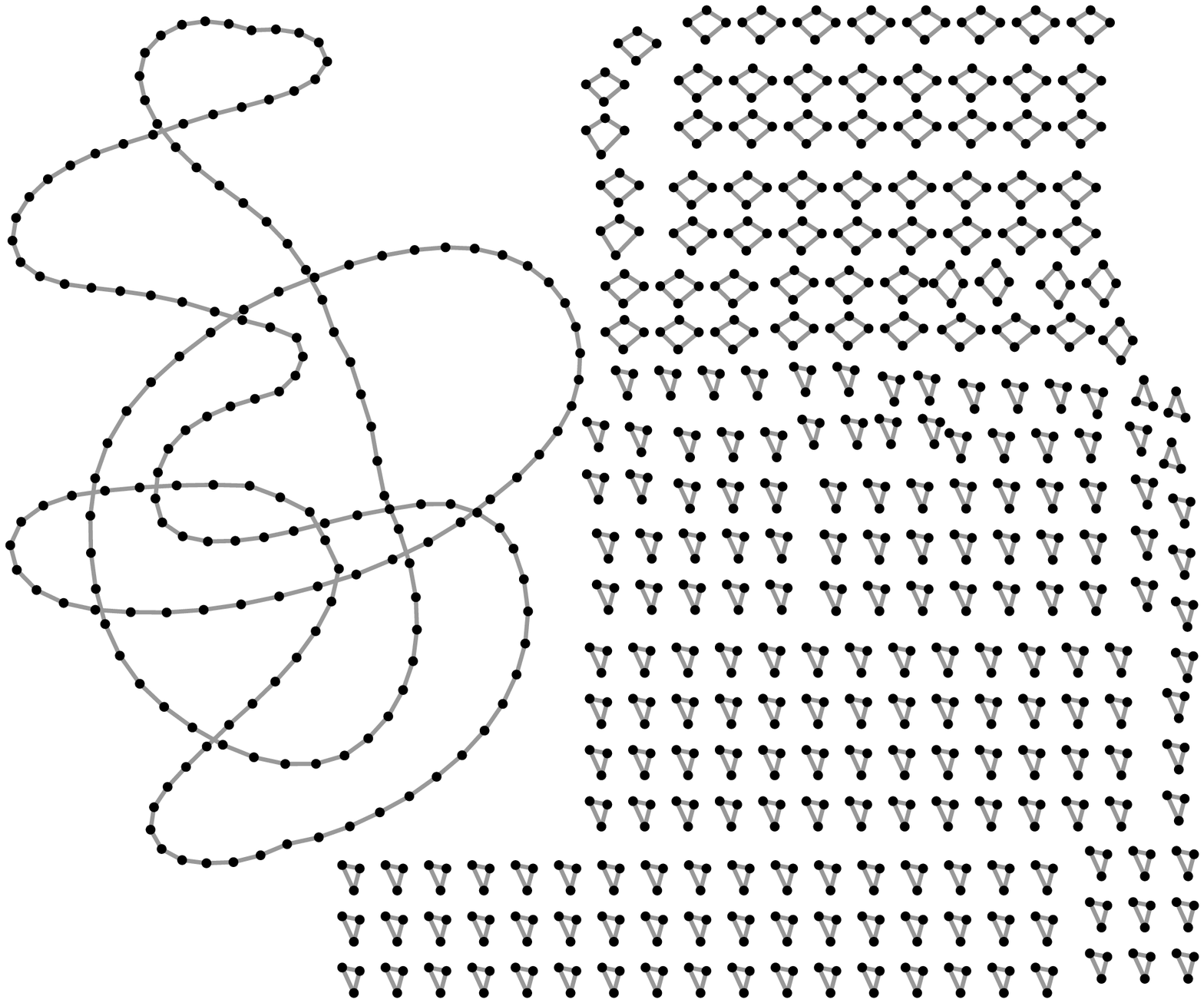}}
  \put(240,140){\includegraphics[width=0.32\textwidth,trim={0 0 0 0},clip]{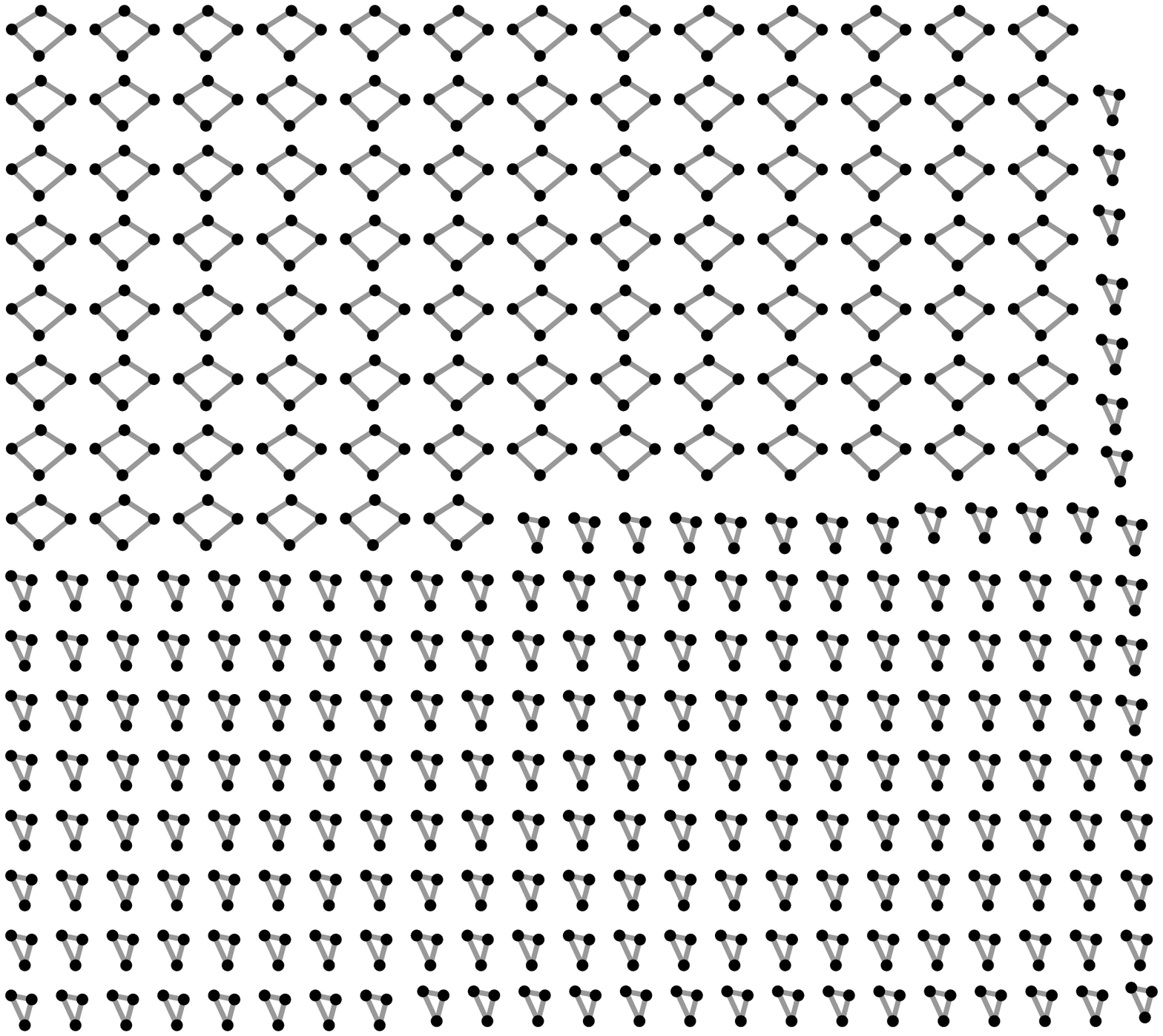}}
  \put(65,6){$\mu$}
  \put(185,6){$\mu$}
  \put(300,6){$\mu$}
  \put(-15,78){$\varrho(\mu)$}
\end{picture}
\vspace*{-6mm}

\caption{\label{fig:graphsAndSpectra} Top row: typical graphs sampled numerically via MCMC from the canonical ensemble (\ref{CD:prob}) of 2-regular nondirected simple graphs, for  $N=1000$.  
Left: $(m_3,m_4)=(0.0,0.06)$ and $m_\infty=0.94$. Middle: $(m_3,m_4)=(0.25,0.56)$  and $m_\infty=0.19$. Right: $(m_3,m_4)=(0.39, 0.61)$  and $m_\infty=0$. 
The bottom row shows the eigenvalue spectra of the corresponding three adjacency matrices, computed by direct numerical  diagonalization.  The locations of the peaks are seen to agree with the theoretical  predictions of  (\ref{CD:spectra}). Note the different scale in the third spectrum graph, to emphasize the weights of the $\delta$-peaks.}
\end{figure*}

For $K=4$  we have two control parameters, $\tilde\alpha_3$ and $\tilde\alpha_4$, 
and the theory claims that for large $N$ the graphs from our ensemble will now be collections of triangles, squares and large rings. Here the key equations (\ref{eq:D_phase},\ref{CD:criticalManifold},\ref{eq:C_phase}) predict that the 
transition line in parameter space is given by $\rme^{3\tilde{\alpha}_3}+\rme^{4\tilde{\alpha}_4}=2$, and that the fractions $m_3$ and $m_4$ of nodes found in triangles and  squares, respectively, are solved (together with the auxiliary order parameter $x$, in the disconnected phase) from:
\begin{eqnarray}
\begin{array}{lll}
\rme^{3\tilde{\alpha}_3}+\rme^{4\tilde{\alpha}_4}<2: &~~ m_3+m_4<1, &~~ {connected~phase},\\[1.5mm]
&~~m_3=\frac{1}{2}\rme^{3\tilde{\alpha}_3},~~m_4=\frac{1}{2}\rme^{4\tilde{\alpha}_4},
\hspace*{-30mm}
\\[4mm] 
\rme^{3\tilde{\alpha}_3}+\rme^{4\tilde{\alpha}_4}>2: &~~ m_3+m_4=1, &~~ {disconnected~phase},\\[1.5mm]
&~~m_3=\frac{1}{2}x^3\rme^{3\tilde{\alpha}_3},~~m_4=\frac{1}{2}x^4\rme^{4\tilde{\alpha}_4}.
\hspace*{-30mm}
\end{array}
\end{eqnarray}
Figure \ref{fig:plotM4} (left panel) shows the resulting predicted phase diagram in the $(\tilde{\alpha}_3,\tilde{\alpha}_4)$ plane.  The mapping $(\tilde\alpha_3,\tilde\alpha_4)\mapsto(m_3,m_4)$ will map the lower region of the phase diagram (the connected phase)  to the interior of the triangle $m_3+m_4<1$ in the right panel of Figure \ref{fig:plotM4}. The upper region of the phase diagram on the left (the disconnected phase), including the critical line, will be mapped to the line $m_3+m_4=1$ in the right panel.
To test also these predictions against numerical simulations, we have chosen multiple points $(\tilde{\alpha}_3,\tilde{\alpha}_4)$ in both regions of the phase diagram, grouped such that the predicted values of $m_\infty=1-m_3-m_4$ were always in the set $\{0.25, 0.5, 0.75\}$. The prediction would therefore be that in the $(m_3,m_4)$ plane these groups of points should be found on the lines $m_3+m_4=1-m_\infty$. Upon measuring the fractions $m_3$ and $m_4$ via MCMC in the corresponding graph ensembles, these predictions are once more validated convincingly.  See Figure \ref{fig:plotM4}. 

Graphs sampled from our ensemble with $K=4$ do indeed typically  consist of controlled numbers of  triangles and   squares, and  a long ring. Figure \ref{fig:graphsAndSpectra} shows examples of such graphs, obtained via MCMC, together with the eigenvalue spectra of their adjacency matrices (obtained by numerical diagonalization).
Also the observed spectra agree with the corresponding theoretical predictions   (\ref{CD:spectra}).

\section{Discussion}

\begin{figure*}[t]
\vspace*{-4mm}
\hspace*{4mm}
\begin{picture}(358,280)
  \put(0,7){\includegraphics[width=0.35\textwidth,height=130pt]{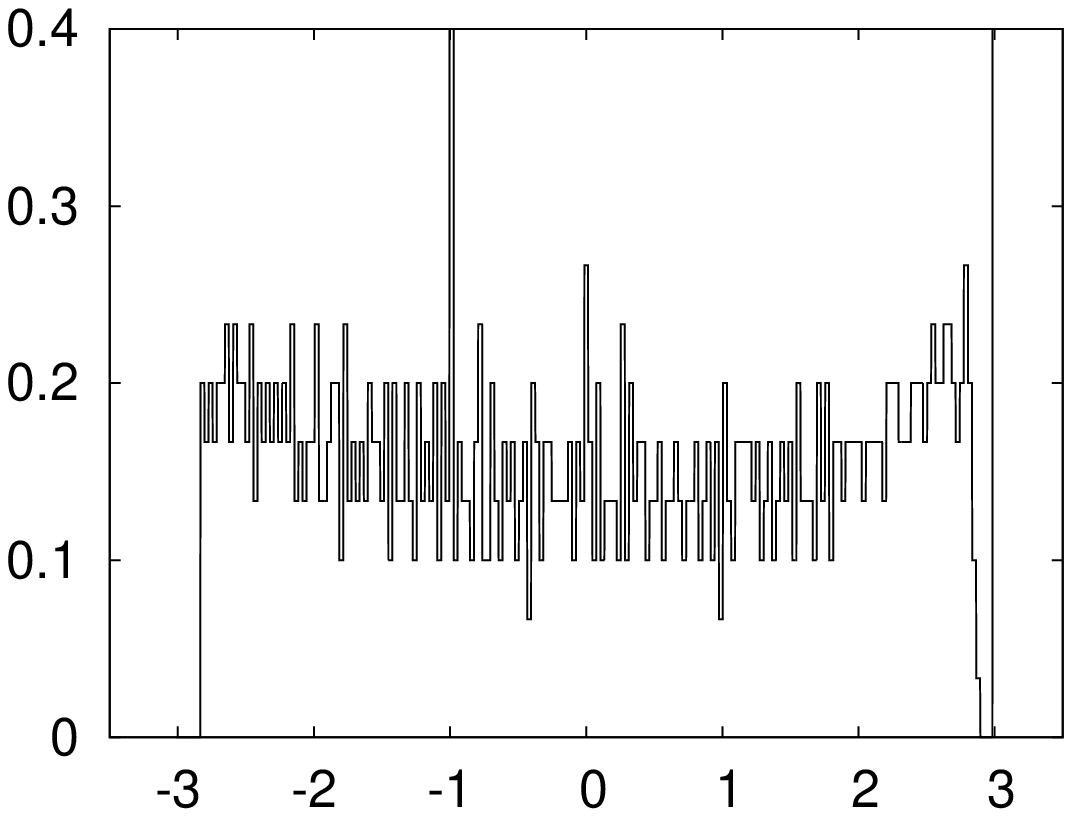}}
  \put(118,7){\includegraphics[width=0.35\textwidth,height=130pt]{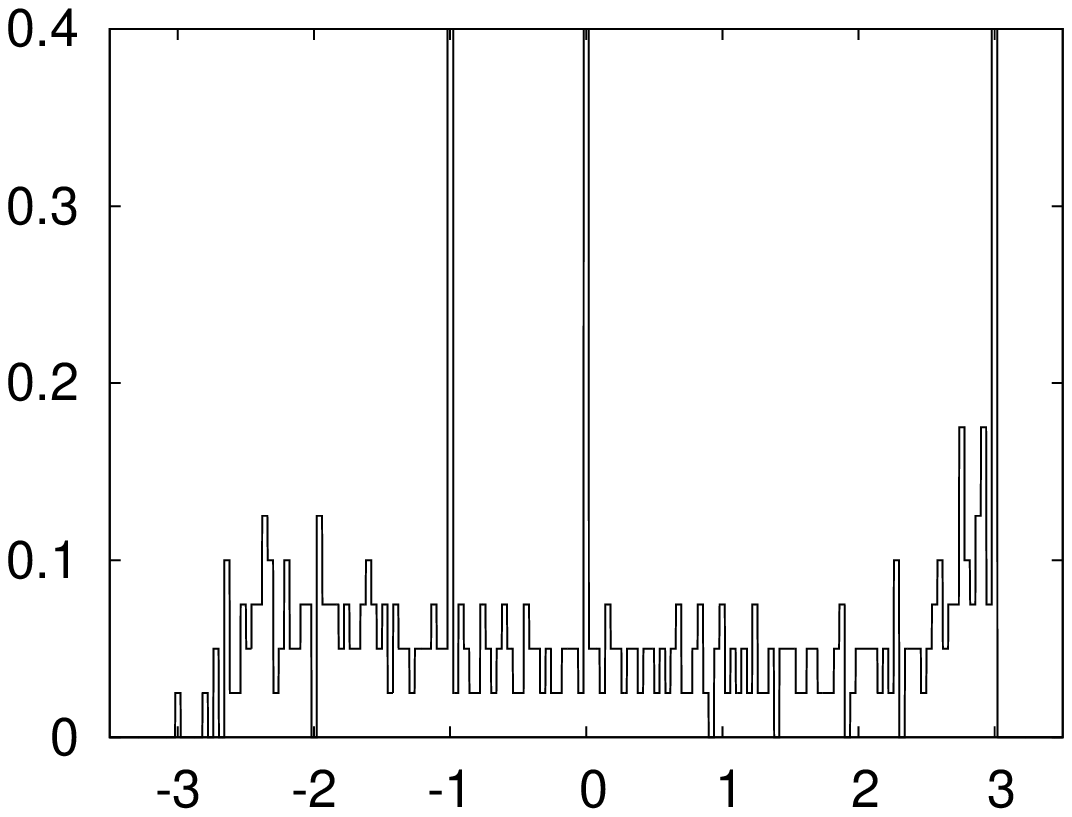}}
  \put(235,7){\includegraphics[width=0.35\textwidth,height=130pt]{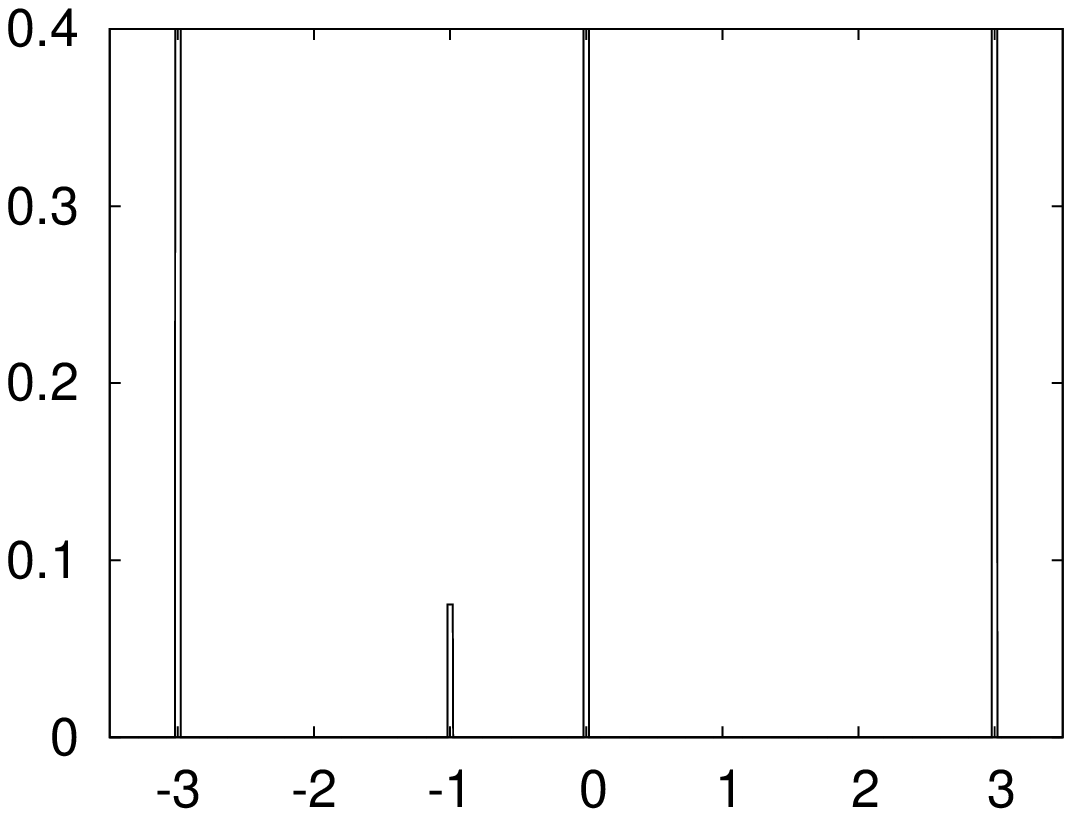}}
  \put(5,140){\includegraphics[width=0.32\textwidth,trim={0 0 0 0},clip]{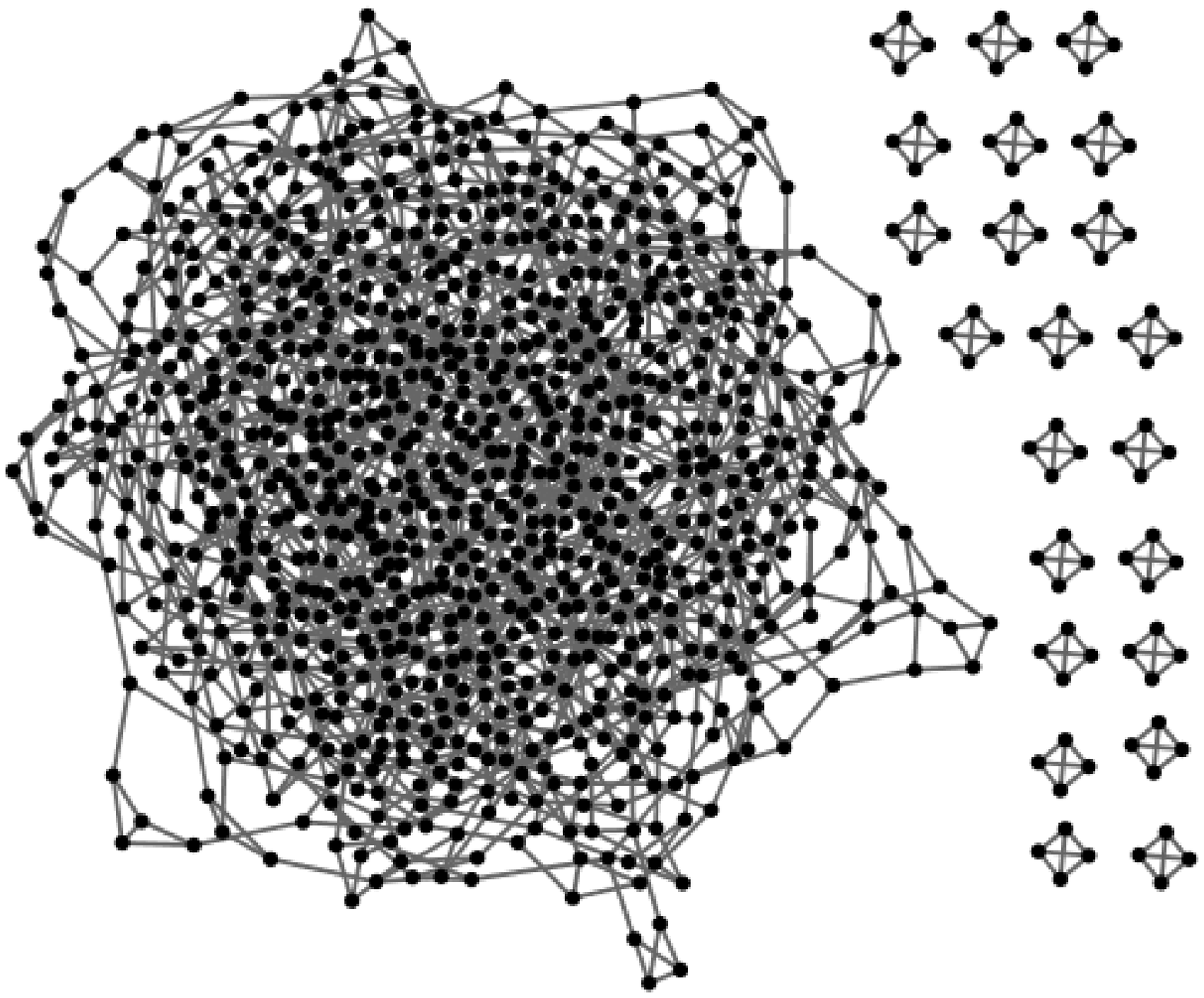}}
  \put(127,135){\includegraphics[width=0.35\textwidth,trim={0 0 0 0},clip]{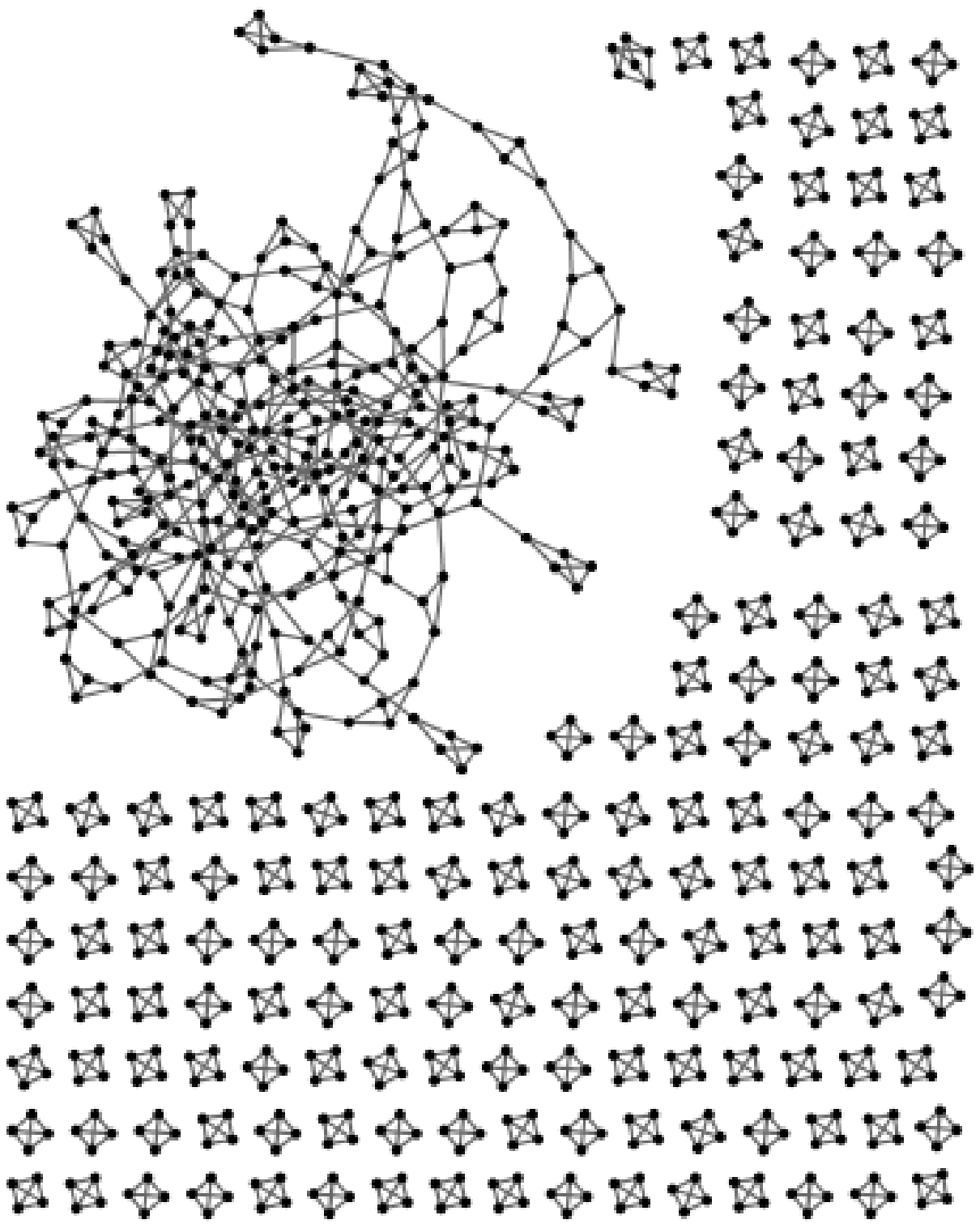}}
  \put(240,140){\includegraphics[width=0.35\textwidth,trim={0 0 0 0},clip]{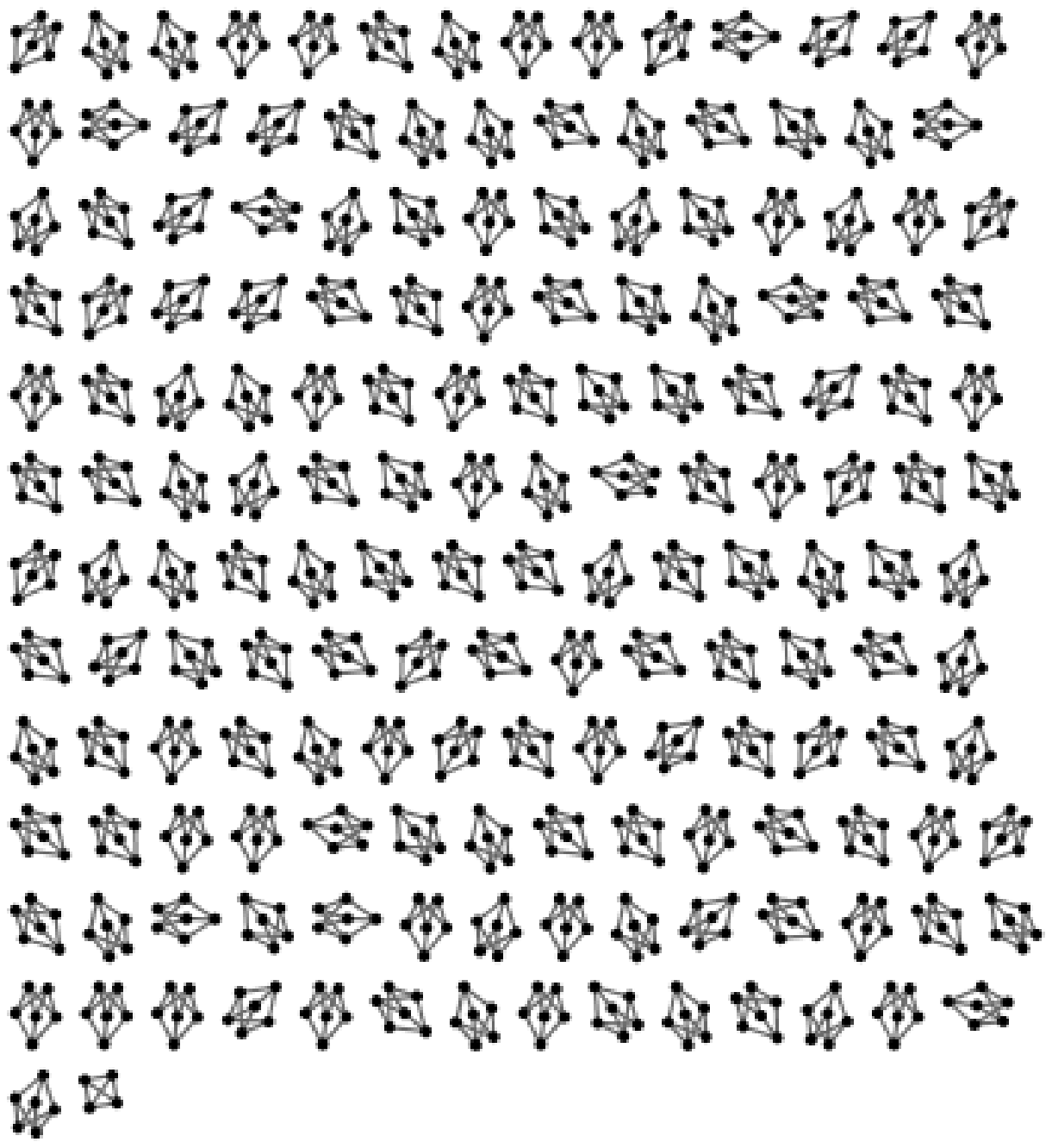}}
  \put(69,6){$\mu$}
  \put(187,6){$\mu$}
  \put(304,6){$\mu$}
  \put(-14,70){$\varrho(\mu)$}
\end{picture}
\vspace*{-6mm}

\caption{\label{fig:graphsAndSpectrak3} Top row: typical graphs sampled numerically via MCMC from the canonical ensemble (\ref{CD:prob}) of 3-regular nondirected simple graphs, for  $N=1000$.  
Left: $(m_3,m_4)=(0.63,0.65)$. Middle: $(m_3,m_4)=(3.34,3.34)$. Right: $(m_3,m_4)=(0.02, 11.97)$. 
The bottom row shows the eigenvalue spectra of the corresponding three adjacency matrices, computed by direct numerical  diagonalization.  The locations of the peaks are seen to agree with the spectrum of the small subgraphs.}
\end{figure*}

In this paper we presented an analytical solution for an exponential random graph ensemble with a controllable density of short cycles. Whereas one would normally not expect such non-treelike graph ensembles to be  solvable, 
here this is possible as a consequence of imposing a local degree constraint of strict 2-regularity. We found  a second order phase transition, which separates a connected phase with large and small cycles from a disconnected phase where the graphs are typically formed only of extensively many short cycles. The short cycles appear in controlled proportions, for which we found analytical expressions in terms of the ensemble's parameters. We also derived an analytical expression for the critical submanifold in the phase diagram, and for the expected eigenvalue spectrum of the graphs' adjacency matrices.  

We analysed both  the canonical and the grand canonical formulation of the ensemble. The canonical version was solved via steepest descent integration. In the grand canonical  version one avoids steepest descent integration, but (as always) the chemical potential takes over the role of the steepest descent integration variable of the canonical version.  In the thermodynamic limit, the canonical and grand canonical routes result in identical equations.  These equations are  found to give highly accurate predictions already for modest graph sizes, such as $N=1000$, as we confirmed in numerical simulations.

The parameter $K$ represents the largest cycle length that is controlled in our model. For $K=3$ one controls only the number of triangles, and  our ensemble becomes similar to that of Strauss \cite{strauss1986general,burda2004network,jonasson1999random} with average degree two. The remaining difference is that in the Strauss model the average degree is imposed implicitly via an overall `soft' constraint, while in the present model all degree values are imposed as local `hard' constraints. 
 Due to this difference,  the degeneration of the Strauss model to a phase where the complete clique has probability one (so the number of triangles can no longer be tuned) is  avoided in the present ensemble. The complete clique is simply no longer an allowed configuration, and hence the number of triangles becomes fully tuneable, if the model parameters scale appropriately with the system size.  In addition, in \cite{yin2016detailed} it is shown that the 'soft' version of our model would have a phase diagram reminiscent of ours. In both cases the sign of a linear combination of functions of the parameters determines the phase of the ensemble. However, in the 'soft' case of \cite{yin2016detailed}  there is a transition from an almost ER-like phase to a clique, while our model exhibits tuneability of the densities in both phases.

 As we mentioned before, the generalization of the present model to other degree distributions has been studied for the Poissonian and the $q$-regular cases. Numerical explorations for 3-regular versions with $K=4$ have been reported in \cite{annibale2017generating}, and show phenomenology similar to that found here for $q=2$.  In particular, one again observes a disconnected phase for large values of $\alpha_3$ and $\alpha_4$.  One could have thought that the phenomenology of our model, in particular the emergence of a large number of small clusters, is specific to the simplifications induced by the 2-regularity condition, but  this is not the case. To emphasize this fact, we redid simulations in the same fashion as in Figure \ref{fig:graphsAndSpectra}, but now for 3-regular as opposed to 2-regular graphs, where analytical solution  along the lines followed for $q=2$  is no longer feasible. 
 The results are shown in Figure \ref{fig:graphsAndSpectrak3}. It is clear that as the bias towards increased numbers of triangles and/or squares is increased, the graph breaks down into small regular graphlets that maximize the cycle density per node.
For Poissonian graphs, simulations revealed in \cite{avetisov2016eigenvalue} that, upon boosting triangles, they also break down into small graphlets, similar to the present model. Also the effect of boosting triangles on the spectrum of the adjacency matrix was similar to what we observed here. However, since the small graphlets that appear in Poissonian models are in general different from isolated triangles and from each other, the associated eigenvalues are described by a different distribution. Similarly, the spectrum of the large component changes in a more complicated way than just by scaling down. Yet, overall we find a similar phenomenology. In fact, our present analysis predicts that the parameters in \cite{avetisov2016eigenvalue} would need a scaling with $N$,  in order  for the transition not to be a finite size effect but to persist in the $N\to\infty$ limit.
 
We could also combine our present model with the Erd\"{o}s-R\`{e}nyi ensemble, to produce connected random graphs with a varying number of short cycles. Again, while the phenomenology of such variations could be explored via simulations, it is not clear how one would be able to obtain analytical solutions without the benefit of 
locally tree-like topology. 

In our view, the main merit of the present model is that its analytical solution helps us understand more complicated `loopy' graph ensembles. We are aware that the analytical route taken in this case is surely impossible for other models. Nevertheless, it provides an explicit analytical solution that reproduces the main features of non-treelike random graph ensembles with hard degree constraints. It helps us understand phenomenology that had so far only been studied numerically. It can also serve as a benchmark model against which more general solution strategies for non-treelike random graphs can be tested, such  as  \cite{CoolenLoopy}, which deals with spectrally constrained  maximum entropy graph ensembles. 
The moments of a graph's spectral density are related to its numbers of cycles,  via the traces of powers of the adjacency matrix. In fact, the present model is a special case of the family of ensembles studied in \cite{CoolenLoopy}, from which it can be obtained by choosing 2-regular degrees and an appropriate polynomial functional Lagrange parameter. 
The analytical and numerical results of this paper suggest that, to obtain phase transitions, the functional Lagrange parameters in spectrally constrained maximum entropy graph ensembles \cite{CoolenLoopy} may need to have a specific scaling with the system size.
\\[3mm]
{\bf Acknowledgement}
\\[1mm]
FAL gratefully acknowledges financial support through a scholarship from Conacyt (Mexico).

\section*{References}

\begin{thebibliography}{10}
\expandafter\ifx\csname url\endcsname\relax
  \def\url#1{{\tt #1}}\fi
\expandafter\ifx\csname urlprefix\endcsname\relax\def\urlprefix{URL }\fi
\providecommand{\eprint}[2][]{\url{#2}}

\bibitem{bekessy1972asymptotic}
B{\'e}k{\'e}ssy A, Bekessy P and Koml{\'o}s J 1972 {\em Stud. Sci. Math.
  Hungar\/} {\bf 7} 343--353

\bibitem{bender1978asymptotic}
Bender E~A and Canfield E~R 1978 {\em Journal of Combinatorial Theory, Series
  A\/} {\bf 24} 296--307

\bibitem{molloy1995critical}
Molloy M and Reed B 1995 {\em Random structures \& algorithms\/} {\bf 6}
  161--180

\bibitem{newman2001random}
Newman M~E, Strogatz S~H and Watts D~J 2001 {\em Physical review E\/} {\bf 64}
  026118

\bibitem{catanzaro2005generation}
Catanzaro M, Bogun{\'a} M and Pastor-Satorras R 2005 {\em Physical Review E\/}
  {\bf 71} 027103

\bibitem{annibale2017generating}
Annibale A, Roberts E {\em et~al.\/} 2017 {\em Generating Random Networks and
  Graphs\/} (Oxford University Press)

\bibitem{holland1983stochastic}
Holland P~W, Laskey K~B and Leinhardt S 1983 {\em Social networks\/} {\bf 5}
  109--137

\bibitem{jonasson1999random}
Jonasson J 1999 {\em Journal of Applied Probability\/} {\bf 36} 852--867

\bibitem{holme2002growing}
Holme P and Kim B~J 2002 {\em Physical review E\/} {\bf 65} 026107

\bibitem{davidsen2002emergence}
Davidsen J, Ebel H and Bornholdt S 2002 {\em Physical Review Letters\/} {\bf
  88} 128701

\bibitem{burda2004network}
Burda Z, Jurkiewicz J and Krzywicki A 2004 {\em Physical Review E\/} {\bf 69}
  026106

\bibitem{krapivsky2005network}
Krapivsky P~L and Redner S 2005 {\em Physical Review E\/} {\bf 71} 036118

\bibitem{newman2009random}
Newman M~E 2009 {\em Physical review letters\/} {\bf 103} 058701

\bibitem{bollobas2011sparse}
Bollob{\'a}s B, Janson S and Riordan O 2011 {\em Random Structures \&
  Algorithms\/} {\bf 38} 269--323

\bibitem{bianconi2014triadic}
Bianconi G, Darst R~K, Iacovacci J and Fortunato S 2014 {\em Physical Review
  E\/} {\bf 90} 042806

\bibitem{granovetter1973strength}
Granovetter M~S 1973 {\em American journal of sociology\/} {\bf 78} 1360--1380

\bibitem{sole2002model}
Sol{\'e} R~V, Pastor-Satorras R, Smith E and Kepler T~B 2002 {\em Advances in
  Complex Systems\/} {\bf 5} 43--54

\bibitem{vazquez2003growing}
V{\'a}zquez A 2003 {\em Physical Review E\/} {\bf 67} 056104

\bibitem{marsili2004rise}
Marsili M, Vega-Redondo F and Slanina F 2004 {\em Proceedings of the National
  Academy of Sciences of the United States of America\/} {\bf 101} 1439--1442

\bibitem{ispolatov2005duplication}
Ispolatov I, Krapivsky P and Yuryev A 2005 {\em Physical review E\/} {\bf 71}
  061911

\bibitem{toivonen2006model}
Toivonen R, Onnela J~P, Saram{\"a}ki J, Hyv{\"o}nen J and Kaski K 2006 {\em
  Physica A: Statistical Mechanics and its Applications\/} {\bf 371} 851--860

\bibitem{jackson2007meeting}
Jackson M~O and Rogers B~W 2007 {\em The American economic review\/} {\bf 97}
  890--915

\bibitem{strauss1986general}
Strauss D 1986 {\em SIAM review\/} {\bf 28} 513--527

\bibitem{chatterjee2013estimating}
Chatterjee S, Diaconis P {\em et~al.\/} 2013 {\em The Annals of Statistics\/}
  {\bf 41} 2428--2461

\bibitem{yin2016detailed}
Yin M 2016 {\em Journal of Statistical Physics\/} {\bf 164} 241--253

\bibitem{avetisov2016eigenvalue}
Avetisov V, Hovhannisyan M, Gorsky A, Nechaev S, Tamm M and Valba O 2016 {\em
  Physical Review E\/} {\bf 94} 062313

\bibitem{CoolenLoopy}
Coolen A 2016 Replica methods for loopy sparse random graphs {\em Journal of
  Physics: Conference Series\/} vol 699 (IOP Publishing) p 012022

\bibitem{infotheory}
Cover T~M and Thomas J~A 2012 {\em Elements of information theory\/} (John
  Wiley \& Sons)

\bibitem{mckay1981expected}
McKay B~D 1981 {\em Linear Algebra and its Applications\/} {\bf 40} 203--216

\bibitem{sampling}
Coolen A, De~Martino A and Annibale A 2009 {\em Journal of Statistical
  Physics\/} {\bf 136} 1035--1067

\end{thebibliography}

\end{document}